\documentclass[aps,prresearch,twocolumn,showpacs,amsmath,amssymb,groupaddress,longbibliography]{revtex4-2}
\usepackage[utf8]{inputenc}
\usepackage{graphicx,graphics}
\usepackage{hyperref}
\hypersetup{
  colorlinks=true,linkcolor=blue,urlcolor=blue,citecolor=blue}
\begin{document}
\title{Ultracold Bose Mixtures with Spin-Dependent Fermion-Mediated Interactions}

%
%

\author{Renyuan Liao}\email{ryliao@fjnu.edu.cn}
\affiliation{Fujian Provincial Key Laboratory for Quantum Manipulation and New Energy Materials, College of Physics and Energy, Fujian Normal University, Fuzhou 350117, China}
\affiliation{Fujian Provincial Collaborative Innovation Center for Advanced High-Field Superconducting Materials and Engineering, Fuzhou, 350117, China}
\date{\today}
\begin{abstract}
   We develop a functional integral formulation for binary Bose-Einstein condensates coupled to polarized fermions. We find that spin-dependent fermion-mediated interactions have dramatic effects on the properties of the binary condensates. The quasiparticle spectrum features two branches. The upper branch, which is of density nature, gets modified by the induced interactions, while the lower branch, which is of spin nature, is left intact. The ground-state phase diagram consists of stable region of miscible phases and unstable region toward phase separation. In the stable region, it is further classified by the damping of excitations of the upper branch. We show that it is possible to find region of well-defined, long-lived quasiparticle excitations by tuning relevant parameters, such as  boson-fermion mass ratio, boson-fermion number density ratio, and interspecies interactions between bosons as well. We explore the effects of quantum fluctuation due to the effective potential on the binary condensates. It turns out that both the density structure factor and spin density structure factor fulfill the Feynman relation, except that the latter is immune to the fermion-mediated interactions.
\end{abstract}
\maketitle

Mediated-interactions are ubiquitous in nature. In high energy physics, all fundamental interactions are mediated by gauge bosons~\cite{WEI95}. In solid state physics, phonon-mediated electron-electron attractions are responsible for the formation of  Cooper pairs, whose condensation leads to the phenomena of conventional superconductivity~\cite{BCS57}. Ultracold atomic gases have emerged as a versatile platform to investigate quantum many-body physics~\cite{BLO08,GEO14,BLO17}. Excitingly, it allows one to create controllable long-range interactions between atoms~\cite{CHI10,ESS13}, including direct electric and magnetic dipole interactions~\cite{MOL10,ZWI15,LAH16,PFA16,GRO17,YE17,FER17}, phonon-mediated coupling in trapped ions~\cite{MON10,BLA11,BOL12}, and photon-mediated interactions in optical cavities~\cite{ESS10,HEM15,DON17,THO18,LEV18,LEV19}. Very recently, adding to the new excitements are the observations of fermion-mediated long-range interactions between bosons in Bose-Fermi mixtures~\cite{CHI19,BRU19,OZE20}. There exist some theoretical efforts~\cite{TIM08,SPI14,BRU15} for understanding such fermion-mediated interactions based on the linear response theory.

 Binary Bose-Einstein condensates (BECs) have been the focus of both theoretical~\cite{SHE96,BIG98,CHU98,TIM98,TIM12,DUI15,PRO16,STR19,WEN20} and experimental~\cite{COR98,WIE08,HIR10,OBE11,SCH13,FER18,PRO18} research over the past years. One of the key questions to ask is how this newly-achieved interaction reshapes our understanding of these exciting systems. Since homogeneous quantum gases have been realized in experiments~\cite{ZWI17,HAD17}, theoretical understanding on the effects of fermion-mediated interactions upon homogeneous binary Bose mixtures becomes an experimentally interesting and urgent task.

 In this work, we shall carry out a systematic study on a homogenous Bose-Fermi mixture, with the aim of laying down a solid and decent framework to treat such problems, fully characterizing fermion-induced interactions, and identifying new features arising from spin-dependent fermion-mediated interactions on binary BECs. First, we shall start from the functional representation of the partition function. By tracing out the fermions, we obtain an effective action entirely in terms of bosonic degrees of freedom, so that we can isolate the effects of fermion-mediated interactions upon the BECs. Second, we shall obtain the induced interactions in the static limit, where an analytic result for the effective interaction potential in real space exists. Third, we will examine how the induce interactions modify the Bogoliubov spectrum and lead to the damping of quasiparticles. Finally, we will explore the quantum fluctuation effects of the effective interaction potential on the density and spin density response of the binary BECs.

We consider a homogeneous mixture of two-species Bose gases and spin-polarized Fermi gases, described by the following grand canonical Hamiltonian
\begin{subequations}
\begin{eqnarray}
   H&=&\int d^3\mathbf{r}\left(\mathcal{H}_B+\mathcal{H}_F+\mathcal{H}_I\right),\\
   \mathcal{H}_B&=&\sum_{\sigma=\uparrow,\downarrow}\phi_\sigma^\dagger \left(-\frac{\hbar^2\nabla^2}{2m_B}-\mu_{\sigma}\right)\phi_\sigma,\\
   \mathcal{H}_F&=&\psi^\dagger (-\frac{\hbar^2\nabla^2}{2m_F}-\mu_F)\psi,\\
   \mathcal{H}_I&=&\sum_{\sigma=\uparrow,\downarrow} \left(g_{F\sigma}\psi^\dagger\psi n_\sigma+\frac{g_{\sigma\sigma}}{2}n_\sigma^2\right)+g_{\uparrow\downarrow}n_\uparrow n_\downarrow.
\end{eqnarray}
\end{subequations}
Here, $\phi_\sigma$ is the field operator for bosons with hyperfine state $\sigma=\{\uparrow,\downarrow\}$, $\mu_{\sigma}$ is the associated chemical potential, and $m_{B}$ is the mass of the atoms. For fermions, $\psi$ is the field operator and $\mu_F$ is the chemical potential. In the interaction term $\mathcal{H}_I$, $n_\sigma=\phi_\sigma^\dagger\phi_\sigma$ is the number operator for bosons of species $\sigma$, the coupling $g_{F\sigma}$ accounts for the interactions between the fermions and the bosons of species $\sigma$, and $g_{\sigma\sigma}$ accounts for the intraspecies interactions between bosons of species $\sigma$, while $g_{\uparrow\downarrow}=4\pi\hbar^2 a_{\uparrow\downarrow}/m_B$ characterizes the interspecies interaction between bosons, where $a_{\uparrow\downarrow}$ is the s-wave interspecies scattering length. For convenience, we define the Fermi momentum $k_F=(6\pi^2n_F)^{1/3}$ with $n_F$ being the number density of Fermi gases, the Fermi velocity $v_F=\hbar k_F/m_F$ and the corresponding Fermi energy $E_F=\hbar^2k_F^2/2m_F$. For brevity, we shall adopt natural units $k_B=\hbar=1$ hereafter.

Within the framework of imaginary-time field integral, the partition function of the system can be cast as $\mathcal{Z}=\int d[\bar{\psi},\psi]d[\phi_\sigma^*,\phi_\sigma]e^{-S}$ with the action given by~\cite{SIM10} $S=\int_0^{\beta} d\tau \left[H+\int d^3\mathbf{r} (\bar{\psi}\partial_\tau\psi+\sum_\sigma\phi_\sigma^*\partial_\tau\phi_\sigma)\right]$, where $\beta=1/k_BT$ is the inverse temperature. To single out the fermion-mediated effects, we carry out the integration over the fermionic degrees of freedom, resulting in an effective action solely in terms of bosonic degrees of freedom $S_{eff}=\int d\tau d^3\mathbf{r}\mathcal{L}_{B}-Tr\ln\mathcal{M}$, where
\begin{eqnarray}
   \!\mathcal{L}_B&=&\sum_\sigma\left[\phi_\sigma^*(\partial_\tau-\frac{ \nabla^2}{2m_B}-\mu_\sigma)\phi_\sigma+\frac{g_{\sigma\sigma}}{2}n_\sigma^2\right]\!+g_{\uparrow\downarrow}n_\uparrow n_\downarrow,\nonumber\\
   \mathcal{M}&=&\partial_\tau-\frac{\nabla^2}{2m_F}-\mu_F+\sum_\sigma g_{F\sigma} \phi_\sigma^*\phi_\sigma.
\end{eqnarray}
Up to this level, the formal manipulation of the partition function is exact. To distill low energy physics, we shall resort to some sorts of approximations to be addressed.

To proceed, we write $\phi_{\sigma}^*\phi_{\sigma}=\rho_{0\sigma}+\sum_{q\neq 0}\rho_{q\sigma}e^{iqx}$ with $x$ being spacetime coordinate and $q\equiv (\mathbf{q},iw_m)$, and set $\mathcal{M}=-\mathcal{G}^{-1}+\mathcal{M}_1$ with $\mathcal{G}^{-1}=-\partial_\tau+\hbar^2\nabla^2/2m_F+\mu_F-\sum_\sigma g_{F\sigma}\rho_{0\sigma}$ being the inverse fermion Green's function and $\mathcal{M}_1=\sum_{q\neq 0}\sum_\sigma g_{F\sigma}\rho_{q\sigma}e^{iqx}$. This allows one to write $Tr\ln\mathcal{M}=Tr\ln(-\mathcal{G}^{-1})+Tr\ln{(1-\mathcal{GM}_1)}$ and to perform series expansions
$-Tr\ln(1-\mathcal{GM}_1)=\sum_{l=1}Tr\left[(\mathcal{G}\mathcal{M}_1)^l\right]/l$. To fully exploit the translational invariance of the system, we shall evaluate the trace in the momentum-frequency representation. We expand the series up to the quadratic order ($l\le2$), resulting in
\begin{subequations}
\begin{eqnarray}
   &&Tr(\mathcal{GM}_1)=\mathcal{M}_1(0) \sum_k \mathcal{G}(k)=0,
   \label{eq:one}\\
   &&\!Tr\left[(\mathcal{GM}_1)^2\right]=\beta V\sum_{q\neq0,\sigma\sigma^\prime}g_{F\sigma}g_{F\sigma^\prime}\Pi_q\rho_{q\sigma}\rho_{-q\sigma^\prime},
   \label{eq:two}\!\\
   &&\Pi_q=\frac{1}{\beta V}\sum_k\mathcal{G}(k)\mathcal{G}(k+q).
   \label{eq:three}
\end{eqnarray}
\end{subequations}
Several comments are in order: For the series expansion, the $l=1$ term vanishes due to $\mathcal{M}_1(0)=0$ by definition, as can be seen from Eq.~$(\ref{eq:one})$; The $l=2$ term corresponds to fermion-induced spin-dependent two-body interactions between bosons, as can be seen from Eq.~$(\ref{eq:two})$ and Eq.~$(\ref{eq:three})$, where we have defined $\Pi_q$, the so-called polarization function; We will neglect $l\geq3$ terms, as they represent induced three-body or more interactions among bosons, which are usually irrelevant for dilute gases. By collecting the relevant terms, we arrive at the approximated effective action $S_{eff}=\int d\tau d\mathbf{r}\mathcal{L}_B-Tr\ln{(-\mathcal{G}^{-1})}+S_{ind}$, with the induced action given by $S_{ind}=\sum_{\sigma\sigma^\prime}g_{F\sigma}g_{F\sigma^\prime}/2\sum_{q\neq 0}\Pi_q\rho_{q\sigma}\rho_{-q\sigma^\prime}$. Remarkably, the induced action is purely a quantum fluctuating effect as the classical $q=0$ component is explicitly excluded.

In the spirit of the Bogoliubov theory, we split the bosonic field $\phi_\sigma$ into a mean-field part $\phi_{0\sigma}$ and a fluctuating part $\varphi_\sigma$. By retaining the fluctuating fields up to the quadratic order, we approximate the effective action as $S_{eff}\approx S_0+S_g$, where $S_0$ is the mean-field action and $S_g$ is the gaussian action with quadratic orders of the fluctuating fields $\varphi_\sigma^*$ and $\varphi_\sigma$. The grand potential density at mean-field level is given by $\Omega^{(0)}=S_0/\beta V$.
The saddle point condition $\delta \Omega^{(0)}/\delta \phi_{0\sigma}^*=0$ leads to the Hugenholz-Pines theorem~\cite{HP59} determining the chemical potential $\mu_\sigma=g_{F\sigma}n_{F\sigma}+g_{\sigma\sigma}\left\vert\phi_{0\sigma}\right\vert^2+g_{\uparrow\downarrow}|\phi_{0\bar{\sigma}}|^2$.
 Without loss of generality, we set $\phi_{0\sigma}=\sqrt{n_{\sigma}}$, where $n_\sigma$ is the condensate density for bosons of species $\sigma$. The self-consistent condition for the fermion density is determined via $n_F=-\partial \Omega^{(0)}/\partial \mu_F$, which gives the chemical potential for the Fermi gases: $\mu_F=E_F+\sum_\sigma g_{F\sigma}n_{\sigma}$.

At the mean-field level, the ground-state energy density can be obtained via $E_G^{(0)}=\Omega^{(0)}+\mu_Fn_F+\sum_\sigma \mu_{\sigma}n_{\sigma}$, yielding
\begin{eqnarray}
   E_G^{(0)}=\frac{3}{5}n_FE_F+\sum_\sigma\left(\frac{g}{2}n_{\sigma}^2+g_{I}n_{\sigma}n_F\right)+g_{\uparrow\downarrow}n_\uparrow n_\downarrow.
\end{eqnarray}
For the system to be stable, we naturally requires that the Hessian matrix constructed for the ground state energy to be positive definite, which leads to an extra constraint for a stable miscible phase
\begin{eqnarray}
    n_F^{1/3}<\frac{(6\pi^2)^{2/3}\left(g_{\uparrow\uparrow}g_{\downarrow\downarrow}-g_{\uparrow\downarrow}^2\right)}{3m_F\left(g_{F\uparrow}^2g_{\downarrow\downarrow}+g_{F\downarrow}^2g_{\uparrow\uparrow}-2g_{F\uparrow}g_{F\downarrow}g_{\uparrow\downarrow}\right)},
    \label{eq:stability}
\end{eqnarray}
in addition to the traditional miscible condition $g_{\uparrow\uparrow}g_{\downarrow\downarrow}-g_{\uparrow\downarrow}^2>0$ where $g_{\sigma\sigma}>0$ for a binary mixtures of Bose-Einstein condensates~\cite{SHE96,BIG98,CHU98,TIM98}.

Derived from $S_{ind}$, the Hamiltonian describing the induced two-body interactions between bosons through coupling with fermions reads  $H_{ind}=\sum_{\sigma\sigma^\prime}g_{F\sigma}g_{F\sigma^\prime}/2\sum_\mathbf{q\neq 0}\Pi(\mathbf{q})\sum_{\mathbf{k,p}}\phi_{\mathbf{k+q}\sigma}^\dagger\phi_{\mathbf{p-q}\sigma^\prime}^\dagger\phi_{\mathbf{p}\sigma^\prime}\phi_{\mathbf{k}\sigma}$.
  Here, $\Pi_\mathbf{q}\equiv\Pi_{(\mathbf{q},0)}$ is the polarization function evaluated at the static  limit at zero temperature, which reads
\begin{eqnarray}
  \Pi_\mathbf{q}=-\frac{d(E_F)}{4}\left[1+\frac{k_F^2-q^2/4}{k_Fq}\ln\left\vert\frac{q+2k_F}{q-2k_F}\right\vert\right],\label{eq:pi}
\end{eqnarray}
where $d(E_F)=m_Fk_F/\pi^2$ is the density of states at the Fermi energy.

Performing the Fourier transform of $H_{ind}$ to real space, we obtain an induced pairwise spin-dependent interaction potential between two Bose atoms of species $\sigma$ and $\sigma^\prime$ with relative coordinate  $\mathbf{r}$, given by
\begin{subequations}
\begin{eqnarray}
   V_{ind}^{\sigma\sigma^\prime}(\mathbf{r})&=&-\frac{d(E_F)g_{F\sigma}g_{F\sigma^\prime}}{4}V_{RKKY}(r),\\
   V_{RKKY}(r)&=&\frac{\sin{(2k_Fr)}-2k_Fr\cos{(2k_Fr)}}{2\pi k_Fr^4}.
\end{eqnarray}
\end{subequations}
The induced spin-dependent attractive long-range interaction is of the RKKY type~\cite{RKKY54} in real space, where it decays at $1/r^3$ at large spatial separation and shows the Friedel oscillations at a period of $1/2k_F$, imprinted by the density of the Fermi gases.

Let us examine how the fermion-mediated interactions plays its role in the cerebrated Bogoliubov theory. For ease of notation, we focus on the situation where both boson species possess same number density and  same intraspecies interaction strength: $n_\uparrow=n_\downarrow\equiv n_B$, $g_{\uparrow\uparrow}=g_{\downarrow\downarrow}\equiv g=4\pi\hbar^2 a_{BB}/m_B$ and $g_{F\uparrow}=g_{F\downarrow}\equiv g_{FB}=2\pi \hbar^2a_{FB}(m_F^{-1}+m_B^{-1})$, where $a_{BB}$ and $a_{FB}$ are the respective s-wave scattering lengths.  By defining a column vector $\Phi_q=(\varphi_{q\uparrow},\varphi_{q\downarrow},\varphi_{-q\uparrow}^\dagger,\varphi_{-q\downarrow}^\dagger)^T$, the gaussian action can be written in a compact form $S_g=(1/2)\sum_q\Phi_q^\dagger \mathcal{G}_B^{-1}\Phi_q$, with the inverse Green's function $\mathcal{G}_B^{-1}(\mathbf{q},i\omega_m)$ defined as follows
\begin{eqnarray}
\mathcal{G}_B^{-1}(\mathbf{q},z)=\begin{pmatrix}
-z+a &  b_1 &  b_2 & b_1\\
b_1&-z+a& b_1 & b_2\\
b_2 & b_1 & z+a & b_1\\
b_1 & b_2 & b_1 & z+a
\end{pmatrix},
\end{eqnarray}
where $a=\epsilon_\mathbf{q}+(g+g_{FB}^2\Pi_q)n_B$ with $\epsilon_\mathbf{q}=\mathbf{q}^2/2m_B$, $b_1=(g_{\uparrow\downarrow}+g_{FB}^2\Pi_q)n_B$, and $b_2=(g+g_{FB}^2\Pi_q)n_B$. The quasiparticle spectrum $\omega(\mathbf{q})$ and the damping rate $\gamma(\mathbf{q})$ can be found by solving the secular equation $det\mathcal{G}_B^{-1}(\mathbf{q},\omega-i\gamma)=0$ with the substitution of $\Pi_q|{i\omega_m\rightarrow \omega+i0^\dagger}$. This substitution corresponds to  analytic continuation to real frequency ($i\omega\rightarrow \omega+i0^\dagger$), yielding the real and imaginary part of the polarization function, so-called the Lindhard function~\cite{LIN54}
\begin{subequations}
\begin{eqnarray}
  Re\Pi&=&-\frac{d(E_F)}{4}\left[1+\sum_{s=\pm}s\frac{1-u_s^2}{2q/k_F}\ln{\left\vert\frac{1+u_s}{1-u_{\bar{s}}}\right\vert}\right],\\
  Im\Pi&=&d(E_F)\frac{\pi k_F}{8q}\left[\sum_{s=\pm}s(1-u_s^2)\Theta(1-u_s^2)\right],\label{eq:imp}
\end{eqnarray}
\end{subequations}
where $u_\pm=\omega/qv_F\pm q/2k_F$ and $\bar{s}=-s$. The system accommodates two branches of excitations $\omega_\pm(\mathbf{q})$, where the upper branch is given by $\omega_+(\mathbf{q})=Re \sqrt{\epsilon_\mathbf{q}\left[\epsilon_\mathbf{q}+2(g+g_{\uparrow\downarrow}+2g_{FB}^2\Pi_q)n_B\right]}$ and the lower branch is given by $\omega_-(\mathbf{q})=\sqrt{\epsilon_\mathbf{q}\left[\epsilon_\mathbf{q}+2(g-g_{\uparrow\downarrow})n_B\right]}$. It is evident that the upper branch gets modified by the induced interaction, while the lower branch is left intact.
\begin{figure}[t]
\includegraphics[width=1.0\columnwidth,height=0.9\columnwidth]{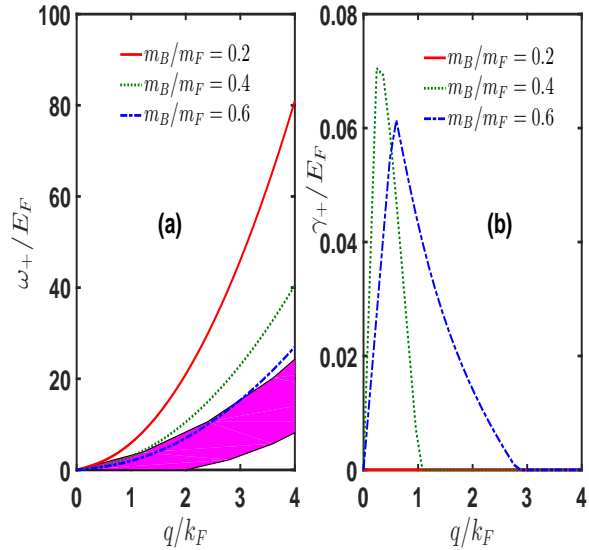}
\caption{(color online) (a) The excitation spectrum for the upper branch $\omega_+(\mathbf{q})$ and (b) the associated damping rate $\gamma_+(\mathbf{q})$ as a function of momentum amplitude $q$ for three typical mass ratios $m_B/m_F=0.2, 0.4, 0.6$. In the shade region in panel (a), the quasiparticle excitation $\omega_+(\mathbf{q})$ is damped with a finite lifetime, whereas outside the region it has infinite lifetime. The parameters we choose are: $k_Fa_{BB}=0.3$, $g_{\uparrow\downarrow}/g=0.6$, and $n_B/n_F=1.0$, which ensures that the system is weakly interacting. }
\label{fig1}
\end{figure}

The quasiparticle spectrum for the upper branch $\omega_+(\mathbf{q})$ and the associated damping rates $\gamma_+(\mathbf{q})$  for three typical mass ratios $m_B/m_F=0.2$, $0.4$, and $0.6$ are shown in Fig.~\ref{fig1}. On panel (a), the shade region has nonzero damping rates. This is found by requiring $Im\Pi\neq 0$, which leads to an inequality constraint $(q/k_F)^2-2q/k_F<\omega/E_F<(q/k_F)^2+2q/k_F$ in the $(\mathbf{q},\omega)$ plane. For a small mass ratio $m_B/m_F=0.2$, the excitation spectrum lies outside the shade region, indicating that it is well-defined and long-lived. With a large mass ratio, the excitation gets damped, propagating with a finite lifetime $1/\gamma$. As shown in panel (b), the damping rate remains zero for varying momentum amplitude at $m_B/m_F=0.2$. For a larger mass ratio, the damping rate features a sharp peak, with the momentum amplitude at which the maximum damping occurs shifts to a higher value.
\begin{figure}[t]
\includegraphics[width=1.0\columnwidth,height=0.9\columnwidth]{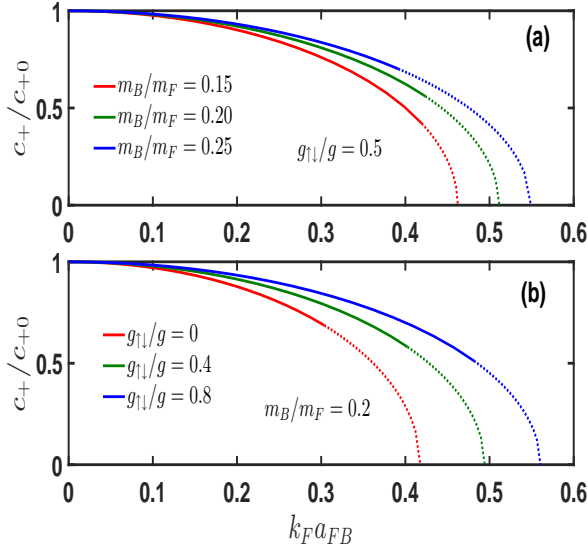}
\caption{(color online) The sound velocity for the upper branch of the excitation spectrum  $c_+/c_{+0}$ (where $c_{+0}=\sqrt{(g+g_{\uparrow\downarrow})n_B/m_B}$) as a function of Bose-Fermi coupling parameter $k_Fa_{FB}$: (a) for three typical mass ratios $m_B/m_F=0.15, 0.20, 0.25$ ; and (b) for three typical interspecies coupling strengths $g_{\uparrow\downarrow}/g=0, 0.4, 0.8$. A solid line denotes that it is an undamped mode while dotted line denotes that the mode is damped. The sound velocity terminates at a critical value of $k_Fa_{FB}$, marking a phase boundary. The parameters we choose are: $k_Fa_{BB}=0.4$ and $n_B/n_F=1.0$. }
\label{fig2}
\end{figure}

At long wavelength, the excitation spectrum is phonon-like with characteristic dispersion $\omega(\mathbf{q})=cq$, where $c$ is the sound velocity. We show the sound velocity $c_+$ for the upper branch of the excitation spectrum  in Fig.~\ref{fig2}. In the absence of Bose-Fermi coupling $k_Fa_{FB}=0$, the sound velocity reduces to $c_{+0}=\sqrt{(g+g_{\uparrow\downarrow})n_B/m_B}$. As the coupling parameter $k_Fa_{FB}$ increases, the sound velocity decreases monotonically, first with a solid line (undamped mode), then with a dotted line (damped mode), before it terminates at zero when reaching a phase boundary separating miscible and immiscible phases. Interestingly, one can verify that the positivity of the sound velocity ($g+g_{\uparrow\downarrow}+2g_{BF}^2\Pi_0>0$) leads to a constraint consistent with Eq.~(\ref{eq:stability}), which can be translated as $(k_Fa_{FB})^2<\pi k_F(a_{BB}+a_{\uparrow\downarrow})m_Fm_B/(m_F+m_B)^2$. Increasing the mass ratio $m_B/m_F$ leads to an increase of the sound velocity, as shown in panel (a). The larger interspecies coupling between bosons $g_{\uparrow\downarrow}/g$, the bigger the sound velocity, as indicated in panel (b).
\begin{figure}[t]
\includegraphics[width=1.0\columnwidth,height=0.9\columnwidth]{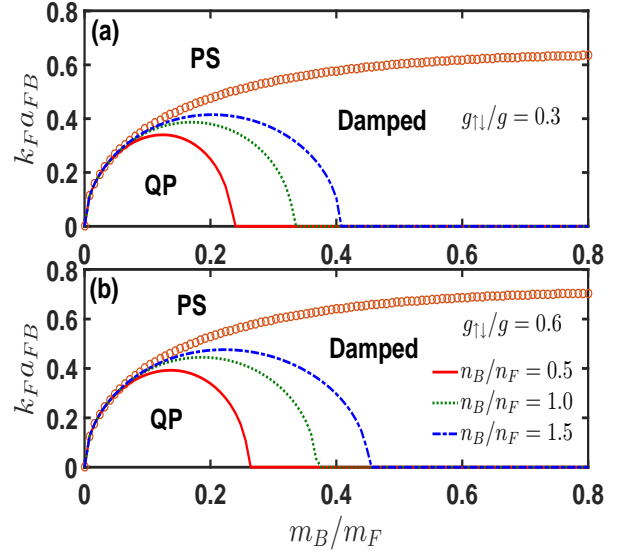}
\caption{(color online) Phase diagram spanned by mass ratio $m_B/m_F$ and Bose-Fermi coupling parameter $k_Fa_{FB}$ for three typical number density ratios $n_B/n_F=0.5, 1.0, 1.5$: (a) $g_{\uparrow\downarrow}/g=0.3$ and (b) $g_{\uparrow\downarrow}/g=0.6$. The phase diagram consists of three regions labeled by PS, QP, and Damped, respectively. PS stands for phase separation, QP stands for quasiparticle with infinite lifetime, and Damped stands for quasiparticle with a finite lifetime. Here we set $k_Fa_{BB}=0.4$, which sets bosons in a weakly-interacting regime. }
\label{fig3}
\end{figure}

We are now in a position to construct a ground-state phase diagram, spanned by mass ratio $m_B/m_F$ and Bose-Fermi coupling parameter $k_Fa_{FB}$. The phase stability constraint provided by Eq.~(\ref{eq:stability}) marks the phase boundary separating stable miscible phase and phase separation (PS) into bosons and fermions or a purely boson phase coexisting with a mixed phase~\cite{MOL98,VIV00,ROT02}, shown in Fig.~\ref{fig3}, which stays intact for varying number density ratio $n_B/n_F$. In the stable miscible phase, we can further classify it into region accommodating long-lived quasiparticle excitations and region residing quasiparticle excitations of finite lifetime due to the Landau damping. To search for well-behaved, long-lived excitations, we consider the region satisfying $Im\Pi(\mathbf{q},\omega)=0$, which occurs at $\omega/qv_F>1+q/2k_F$ [see Eq.~(\ref{eq:imp})]. At long wavelength, this becomes $c_+/E_F>2k_F$, yielding
$(k_Fa_{FB})^2<\pi m_Fm_B/(m_F+m_B)^2\left[k_F(a_{BB}+a_{\uparrow\downarrow})-3\pi n_Fm_B^2/(2n_Bm_F^2)\right]$. As shown in Fig.~\ref{fig3}, tuning up the number density ratio $n_B/n_F$ expands the region of quasiparticle excitations with infinite lifetime (QP). Increasing the interspecies coupling $g_{\uparrow\downarrow}/g$ contributes to a broadened region of QP, and a shrinkage of damped region. The reason behind the fact that we have focused on the Landau damping of the collective long-wavelength excitations is that Beliaev damping is strongly suppressed at low momenta~\cite{LIU97}.

Now we turn to examine the effects of quantum fluctuation arising from the effective interaction potential on the properties of the binary BECs.
The static structure factor $S(\mathbf{q})$ probes density fluctuations of a system. It provides information both on the spectrum of collective excitations and the momentum distribution. We can evaluate the static structure at the Bogoliubov level as follows:
\begin{eqnarray}
  S(\mathbf{q})&=&\frac{1}{2N_B}<\delta\rho_{\mathbf{q}}^\dagger\delta\rho_\mathbf{q}>=\frac{1}{2}\sum_{i\omega_m}\sum_{i,j=1}^4\mathcal{G}_{Bij}(\mathbf{q},i\omega_m)\nonumber\\
  &=&\frac{\epsilon_\mathbf{q}}{\omega_+(\mathbf{q})}\cot{\frac{\beta\omega_+(\mathbf{q})}{2}}.
\end{eqnarray}
This corresponds to the Feynman relation~\cite{FEY54} for the upper branch, which connects the static structure factor to the excitation spectrum for a Bose superfluid enjoys time-reversal symmetry. The Bragg spectroscopy can be employed to measure the zero-temperature structure factor of the system~\cite{KET99}.

Similarly, the spin density structure factor $S_\sigma(\mathbf{q})$ can be determined as follows:
\begin{eqnarray}
   S_\sigma(\mathbf{q})&=&\frac{1}{2N_B}<\delta(\rho_{\mathbf{q}\uparrow}-\rho_{\mathbf{q}\downarrow})^\dagger\delta(\rho_{\mathbf{q}\uparrow}-\rho_{\mathbf{q}\downarrow})>\nonumber\\
   &=&\frac{1}{2}\sum_{i\omega_m}\sum_{i,j=1}^4(-1)^{i+j}\mathcal{G}_{Bij}(\mathbf{q},i\omega_m)\nonumber\\
   &=&\frac{\epsilon_\mathbf{q}}{\omega_-(\mathbf{q})}\cot{\frac{\beta\omega_-(\mathbf{q})}{2}}.
\end{eqnarray}
This is precisely the Feynman relation for the lower branch, which connects the spin structure factor to the excitation spectrum of the Bose superfluid. Since the lower branch is left intact for the induced interaction, we deduce that the spin density structure factor is immune to the Bose-Fermi coupling. The spin structure factor can be measured from noise correlations or Bragg scattering of light~\cite{HUL15,GRE17}.

In summary, we find that spin-dependent fermion-mediated interactions dramatically modify the properties of the binary BECs. The upper branch is affected by the induced interactions, while the lower branch is clearly not. We map out the phase diagram based on the phase stability condition and Landau damping of the excitations of the upper branch. It consists of the phase boundary separating stable region of miscible phases and unstable region toward phase separation. Due to the fermion-mediated interactions, we find that the stable region can be further classified into two parts based on the damping of the excitations. The predicted damping rate can be probed experimentally via two-phonon Bragg spectroscopy~\cite{DAV05}. We find that both the density structure factor and spin density structure factor satisfy the Feynman relation, reflecting density and spin excitations respectively. Experimental verifications of new features predicted in this work is expected to provide a significant advance to our understanding of emergent phenomena associated with spin-dependent fermion-mediated interactions.
\section*{acknowledgments}
This work is supported by NSFC under Grant No.11674058 and NCET-13-0734.

\begin{thebibliography}{60}%
\makeatletter
\providecommand \@ifxundefined [1]{%
 \@ifx{#1\undefined}
}%
\providecommand \@ifnum [1]{%
 \ifnum #1\expandafter \@firstoftwo
 \else \expandafter \@secondoftwo
 \fi
}%
\providecommand \@ifx [1]{%
 \ifx #1\expandafter \@firstoftwo
 \else \expandafter \@secondoftwo
 \fi
}%
\providecommand \natexlab [1]{#1}%
\providecommand \enquote  [1]{``#1''}%
\providecommand \bibnamefont  [1]{#1}%
\providecommand \bibfnamefont [1]{#1}%
\providecommand \citenamefont [1]{#1}%
\providecommand \href@noop [0]{\@secondoftwo}%
\providecommand \href [0]{\begingroup \@sanitize@url \@href}%
\providecommand \@href[1]{\@@startlink{#1}\@@href}%
\providecommand \@@href[1]{\endgroup#1\@@endlink}%
\providecommand \@sanitize@url [0]{\catcode `\\12\catcode `\$12\catcode
  `\&12\catcode `\#12\catcode `\^12\catcode `\_12\catcode `\%12\relax}%
\providecommand \@@startlink[1]{}%
\providecommand \@@endlink[0]{}%
\providecommand \url  [0]{\begingroup\@sanitize@url \@url }%
\providecommand \@url [1]{\endgroup\@href {#1}{\urlprefix }}%
\providecommand \urlprefix  [0]{URL }%
\providecommand \Eprint [0]{\href }%
\providecommand \doibase [0]{https://doi.org/}%
\providecommand \selectlanguage [0]{\@gobble}%
\providecommand \bibinfo  [0]{\@secondoftwo}%
\providecommand \bibfield  [0]{\@secondoftwo}%
\providecommand \translation [1]{[#1]}%
\providecommand \BibitemOpen [0]{}%
\providecommand \bibitemStop [0]{}%
\providecommand \bibitemNoStop [0]{.\EOS\space}%
\providecommand \EOS [0]{\spacefactor3000\relax}%
\providecommand \BibitemShut  [1]{\csname bibitem#1\endcsname}%
\let\auto@bib@innerbib\@empty
\bibitem [{\citenamefont {Weinberg}(1995)}]{WEI95}%
  \BibitemOpen
  \bibfield  {author} {\bibinfo {author} {\bibfnamefont {S.}~\bibnamefont
  {Weinberg}},\ }\href@noop {} {\emph {\bibinfo {title} {The Quantum Theory of
  Fields}}}\ (\bibinfo  {publisher} {CUP},\ \bibinfo {address} {Cambridge,
  UK},\ \bibinfo {year} {1995})\BibitemShut {NoStop}%
\bibitem [{\citenamefont {Bardeen}\ \emph {et~al.}(1957)\citenamefont
  {Bardeen}, \citenamefont {Cooper},\ and\ \citenamefont {Schrieffer}}]{BCS57}%
  \BibitemOpen
  \bibfield  {author} {\bibinfo {author} {\bibfnamefont {J.}~\bibnamefont
  {Bardeen}}, \bibinfo {author} {\bibfnamefont {L.~N.}\ \bibnamefont
  {Cooper}},\ and\ \bibinfo {author} {\bibfnamefont {J.~R.}\ \bibnamefont
  {Schrieffer}},\ }\bibfield  {title} {\bibinfo {title} {Theory of
  superconductivity},\ }\href@noop {} {\bibfield  {journal} {\bibinfo
  {journal} {Phys. Rev.}\ }\textbf {\bibinfo {volume} {108}},\ \bibinfo {pages}
  {1175} (\bibinfo {year} {1957})}\BibitemShut {NoStop}%
\bibitem [{\citenamefont {Bloch}\ \emph {et~al.}(2008)\citenamefont {Bloch},
  \citenamefont {Dalibard},\ and\ \citenamefont {Zwerger}}]{BLO08}%
  \BibitemOpen
  \bibfield  {author} {\bibinfo {author} {\bibfnamefont {I.}~\bibnamefont
  {Bloch}}, \bibinfo {author} {\bibfnamefont {J.}~\bibnamefont {Dalibard}},\
  and\ \bibinfo {author} {\bibfnamefont {W.}~\bibnamefont {Zwerger}},\
  }\bibfield  {title} {\bibinfo {title} {Many-body physics with ultracold
  gases},\ }\href@noop {} {\bibfield  {journal} {\bibinfo  {journal} {Rev. Mod.
  Phys.}\ }\textbf {\bibinfo {volume} {80}},\ \bibinfo {pages} {885} (\bibinfo
  {year} {2008})}\BibitemShut {NoStop}%
\bibitem [{\citenamefont {Georgescu}\ \emph {et~al.}(2014)\citenamefont
  {Georgescu}, \citenamefont {Ashhab},\ and\ \citenamefont {Nori}}]{GEO14}%
  \BibitemOpen
  \bibfield  {author} {\bibinfo {author} {\bibfnamefont {I.~M.}\ \bibnamefont
  {Georgescu}}, \bibinfo {author} {\bibfnamefont {S.}~\bibnamefont {Ashhab}},\
  and\ \bibinfo {author} {\bibfnamefont {F.}~\bibnamefont {Nori}},\ }\bibfield
  {title} {\bibinfo {title} {Quantum simulation},\ }\href@noop {} {\bibfield
  {journal} {\bibinfo  {journal} {Rev. Mod. Phys.}\ }\textbf {\bibinfo {volume}
  {86}},\ \bibinfo {pages} {153} (\bibinfo {year} {2014})}\BibitemShut
  {NoStop}%
\bibitem [{\citenamefont {Gross}\ and\ \citenamefont {Bloch}(2017)}]{BLO17}%
  \BibitemOpen
  \bibfield  {author} {\bibinfo {author} {\bibfnamefont {C.}~\bibnamefont
  {Gross}}\ and\ \bibinfo {author} {\bibfnamefont {I.}~\bibnamefont {Bloch}},\
  }\bibfield  {title} {\bibinfo {title} {Quantum simulation with ultracold
  atoms in optical lattcices},\ }\href@noop {} {\bibfield  {journal} {\bibinfo
  {journal} {Science}\ }\textbf {\bibinfo {volume} {357}},\ \bibinfo {pages}
  {995} (\bibinfo {year} {2017})}\BibitemShut {NoStop}%
\bibitem [{\citenamefont {Chin}\ \emph {et~al.}(2010)\citenamefont {Chin},
  \citenamefont {Grimm}, \citenamefont {Julienne},\ and\ \citenamefont
  {Tiesinga}}]{CHI10}%
  \BibitemOpen
  \bibfield  {author} {\bibinfo {author} {\bibfnamefont {C.}~\bibnamefont
  {Chin}}, \bibinfo {author} {\bibfnamefont {R.}~\bibnamefont {Grimm}},
  \bibinfo {author} {\bibfnamefont {P.}~\bibnamefont {Julienne}},\ and\
  \bibinfo {author} {\bibfnamefont {E.}~\bibnamefont {Tiesinga}},\ }\bibfield
  {title} {\bibinfo {title} {Feshbach resonances in ultracold gases},\
  }\href@noop {} {\bibfield  {journal} {\bibinfo  {journal} {Rev. Mod. Phys.}\
  }\textbf {\bibinfo {volume} {82}},\ \bibinfo {pages} {1225} (\bibinfo {year}
  {2010})}\BibitemShut {NoStop}%
\bibitem [{\citenamefont {Ritsch}\ \emph {et~al.}(2013)\citenamefont {Ritsch},
  \citenamefont {Domokos}, \citenamefont {Brennecke},\ and\ \citenamefont
  {Esslinger}}]{ESS13}%
  \BibitemOpen
  \bibfield  {author} {\bibinfo {author} {\bibfnamefont {H.}~\bibnamefont
  {Ritsch}}, \bibinfo {author} {\bibfnamefont {P.}~\bibnamefont {Domokos}},
  \bibinfo {author} {\bibfnamefont {F.}~\bibnamefont {Brennecke}},\ and\
  \bibinfo {author} {\bibfnamefont {T.}~\bibnamefont {Esslinger}},\ }\bibfield
  {title} {\bibinfo {title} {Cold atoms in cavity-generated dynamical optical
  potentials},\ }\href {https://doi.org/10.1103/RevModPhys.85.553} {\bibfield
  {journal} {\bibinfo  {journal} {Rev. Mod. Phys.}\ }\textbf {\bibinfo {volume}
  {85}},\ \bibinfo {pages} {553} (\bibinfo {year} {2013})}\BibitemShut
  {NoStop}%
\bibitem [{\citenamefont {Saffman}\ \emph {et~al.}(2010)\citenamefont
  {Saffman}, \citenamefont {Walker},\ and\ \citenamefont {Molmer}}]{MOL10}%
  \BibitemOpen
  \bibfield  {author} {\bibinfo {author} {\bibfnamefont {M.}~\bibnamefont
  {Saffman}}, \bibinfo {author} {\bibfnamefont {T.~G.}\ \bibnamefont
  {Walker}},\ and\ \bibinfo {author} {\bibfnamefont {K.}~\bibnamefont
  {Molmer}},\ }\bibfield  {title} {\bibinfo {title} {Quantum information with
  rydberg atoms},\ }\href@noop {} {\bibfield  {journal} {\bibinfo  {journal}
  {Rev. Mod. Phys.}\ }\textbf {\bibinfo {volume} {82}},\ \bibinfo {pages}
  {2313} (\bibinfo {year} {2010})}\BibitemShut {NoStop}%
\bibitem [{\citenamefont {Park}\ \emph {et~al.}(2015)\citenamefont {Park},
  \citenamefont {Will},\ and\ \citenamefont {Zwierlein}}]{ZWI15}%
  \BibitemOpen
  \bibfield  {author} {\bibinfo {author} {\bibfnamefont {J.~W.}\ \bibnamefont
  {Park}}, \bibinfo {author} {\bibfnamefont {S.~A.}\ \bibnamefont {Will}},\
  and\ \bibinfo {author} {\bibfnamefont {M.~W.}\ \bibnamefont {Zwierlein}},\
  }\bibfield  {title} {\bibinfo {title} {Ultracold dipolar gas of fermionic nak
  molecules in their absolute ground state},\ }\href@noop {} {\bibfield
  {journal} {\bibinfo  {journal} {Phys. Rev. Lett.}\ }\textbf {\bibinfo
  {volume} {114}},\ \bibinfo {pages} {205302} (\bibinfo {year}
  {2015})}\BibitemShut {NoStop}%
\bibitem [{\citenamefont {Labuhn}\ \emph {et~al.}(2016)\citenamefont {Labuhn},
  \citenamefont {Barredo}, \citenamefont {Ravets}, \citenamefont {Leseleuc},
  \citenamefont {Macri}, \citenamefont {Lahaye},\ and\ \citenamefont
  {Browaeys}}]{LAH16}%
  \BibitemOpen
  \bibfield  {author} {\bibinfo {author} {\bibfnamefont {H.}~\bibnamefont
  {Labuhn}}, \bibinfo {author} {\bibfnamefont {D.}~\bibnamefont {Barredo}},
  \bibinfo {author} {\bibfnamefont {S.}~\bibnamefont {Ravets}}, \bibinfo
  {author} {\bibfnamefont {S.}~\bibnamefont {Leseleuc}}, \bibinfo {author}
  {\bibfnamefont {T.}~\bibnamefont {Macri}}, \bibinfo {author} {\bibfnamefont
  {T.}~\bibnamefont {Lahaye}},\ and\ \bibinfo {author} {\bibfnamefont
  {A.}~\bibnamefont {Browaeys}},\ }\bibfield  {title} {\bibinfo {title}
  {Tunable two-dimensional arrays of single ryderberg atoms for realizing
  quantum ising models},\ }\href@noop {} {\bibfield  {journal} {\bibinfo
  {journal} {Nature}\ }\textbf {\bibinfo {volume} {534}},\ \bibinfo {pages}
  {677} (\bibinfo {year} {2016})}\BibitemShut {NoStop}%
\bibitem [{\citenamefont {Kadau}\ \emph {et~al.}(2016)\citenamefont {Kadau},
  \citenamefont {Schmitt}, \citenamefont {Wenzel}, \citenamefont {Wink},
  \citenamefont {Maier}, \citenamefont {Ferrier-Barbut},\ and\ \citenamefont
  {Pfau}}]{PFA16}%
  \BibitemOpen
  \bibfield  {author} {\bibinfo {author} {\bibfnamefont {H.}~\bibnamefont
  {Kadau}}, \bibinfo {author} {\bibfnamefont {M.}~\bibnamefont {Schmitt}},
  \bibinfo {author} {\bibfnamefont {M.}~\bibnamefont {Wenzel}}, \bibinfo
  {author} {\bibfnamefont {C.}~\bibnamefont {Wink}}, \bibinfo {author}
  {\bibfnamefont {T.}~\bibnamefont {Maier}}, \bibinfo {author} {\bibfnamefont
  {I.}~\bibnamefont {Ferrier-Barbut}},\ and\ \bibinfo {author} {\bibfnamefont
  {T.}~\bibnamefont {Pfau}},\ }\bibfield  {title} {\bibinfo {title} {Observing
  the rosensweig instability of a quantum ferrofluid},\ }\href@noop {}
  {\bibfield  {journal} {\bibinfo  {journal} {Nature}\ }\textbf {\bibinfo
  {volume} {530}},\ \bibinfo {pages} {194} (\bibinfo {year}
  {2016})}\BibitemShut {NoStop}%
\bibitem [{\citenamefont {Zeiher}\ \emph {et~al.}(2017)\citenamefont {Zeiher},
  \citenamefont {Choi}, \citenamefont {Rubio-Abadal}, \citenamefont {Pohl},
  \citenamefont {vanBijnen}, \citenamefont {Bloch},\ and\ \citenamefont
  {Gross}}]{GRO17}%
  \BibitemOpen
  \bibfield  {author} {\bibinfo {author} {\bibfnamefont {J.}~\bibnamefont
  {Zeiher}}, \bibinfo {author} {\bibfnamefont {J.-Y.}\ \bibnamefont {Choi}},
  \bibinfo {author} {\bibfnamefont {A.}~\bibnamefont {Rubio-Abadal}}, \bibinfo
  {author} {\bibfnamefont {T.}~\bibnamefont {Pohl}}, \bibinfo {author}
  {\bibfnamefont {R.}~\bibnamefont {vanBijnen}}, \bibinfo {author}
  {\bibfnamefont {I.}~\bibnamefont {Bloch}},\ and\ \bibinfo {author}
  {\bibfnamefont {C.}~\bibnamefont {Gross}},\ }\bibfield  {title} {\bibinfo
  {title} {Coherent many-body spin dynamics in a long-range interacting ising
  chain},\ }\href@noop {} {\bibfield  {journal} {\bibinfo  {journal} {Phys.
  Rev. X}\ }\textbf {\bibinfo {volume} {7}},\ \bibinfo {pages} {041063}
  (\bibinfo {year} {2017})}\BibitemShut {NoStop}%
\bibitem [{\citenamefont {Moses}\ \emph {et~al.}(2017)\citenamefont {Moses},
  \citenamefont {Covey}, \citenamefont {Miecnikowski}, \citenamefont {Jin},\
  and\ \citenamefont {Ye}}]{YE17}%
  \BibitemOpen
  \bibfield  {author} {\bibinfo {author} {\bibfnamefont {S.~A.}\ \bibnamefont
  {Moses}}, \bibinfo {author} {\bibfnamefont {J.~P.}\ \bibnamefont {Covey}},
  \bibinfo {author} {\bibfnamefont {T.~M.}\ \bibnamefont {Miecnikowski}},
  \bibinfo {author} {\bibfnamefont {D.~S.}\ \bibnamefont {Jin}},\ and\ \bibinfo
  {author} {\bibfnamefont {J.}~\bibnamefont {Ye}},\ }\bibfield  {title}
  {\bibinfo {title} {New frontiers for quantum gases of polar molecules},\
  }\href@noop {} {\bibfield  {journal} {\bibinfo  {journal} {Nat. Phys.}\
  }\textbf {\bibinfo {volume} {13}},\ \bibinfo {pages} {13} (\bibinfo {year}
  {2017})}\BibitemShut {NoStop}%
\bibitem [{\citenamefont {Baier}\ \emph {et~al.}(2017)\citenamefont {Baier},
  \citenamefont {Mark}, \citenamefont {Petter}, \citenamefont {Aikawa},
  \citenamefont {Chomaz}, \citenamefont {Cai}, \citenamefont {Baranov},
  \citenamefont {Zoller},\ and\ \citenamefont {Ferlaino}}]{FER17}%
  \BibitemOpen
  \bibfield  {author} {\bibinfo {author} {\bibfnamefont {S.}~\bibnamefont
  {Baier}}, \bibinfo {author} {\bibfnamefont {M.~J.}\ \bibnamefont {Mark}},
  \bibinfo {author} {\bibfnamefont {D.}~\bibnamefont {Petter}}, \bibinfo
  {author} {\bibfnamefont {K.}~\bibnamefont {Aikawa}}, \bibinfo {author}
  {\bibfnamefont {L.}~\bibnamefont {Chomaz}}, \bibinfo {author} {\bibfnamefont
  {Z.}~\bibnamefont {Cai}}, \bibinfo {author} {\bibfnamefont {M.}~\bibnamefont
  {Baranov}}, \bibinfo {author} {\bibfnamefont {P.}~\bibnamefont {Zoller}},\
  and\ \bibinfo {author} {\bibfnamefont {F.}~\bibnamefont {Ferlaino}},\
  }\bibfield  {title} {\bibinfo {title} {Extended bose-hubbard models with
  ultracold magnetic atoms},\ }\href@noop {} {\bibfield  {journal} {\bibinfo
  {journal} {Science}\ }\textbf {\bibinfo {volume} {352}},\ \bibinfo {pages}
  {201} (\bibinfo {year} {2017})}\BibitemShut {NoStop}%
\bibitem [{\citenamefont {Kim}\ \emph {et~al.}(2010)\citenamefont {Kim},
  \citenamefont {Chang}, \citenamefont {Korenblit}, \citenamefont {Islam},
  \citenamefont {Edwards}, \citenamefont {Freericks}, \citenamefont {Lin},
  \citenamefont {Duan},\ and\ \citenamefont {Monroe}}]{MON10}%
  \BibitemOpen
  \bibfield  {author} {\bibinfo {author} {\bibfnamefont {K.}~\bibnamefont
  {Kim}}, \bibinfo {author} {\bibfnamefont {M.-S.}\ \bibnamefont {Chang}},
  \bibinfo {author} {\bibfnamefont {S.}~\bibnamefont {Korenblit}}, \bibinfo
  {author} {\bibfnamefont {R.}~\bibnamefont {Islam}}, \bibinfo {author}
  {\bibfnamefont {E.~E.}\ \bibnamefont {Edwards}}, \bibinfo {author}
  {\bibfnamefont {J.~K.}\ \bibnamefont {Freericks}}, \bibinfo {author}
  {\bibfnamefont {G.-D.}\ \bibnamefont {Lin}}, \bibinfo {author} {\bibfnamefont
  {L.-M.}\ \bibnamefont {Duan}},\ and\ \bibinfo {author} {\bibfnamefont
  {C.}~\bibnamefont {Monroe}},\ }\bibfield  {title} {\bibinfo {title} {Quantum
  simulation of frusrtated ising spins with trapped ions},\ }\href@noop {}
  {\bibfield  {journal} {\bibinfo  {journal} {Nature}\ }\textbf {\bibinfo
  {volume} {465}},\ \bibinfo {pages} {590} (\bibinfo {year}
  {2010})}\BibitemShut {NoStop}%
\bibitem [{\citenamefont {Barreiro}\ \emph {et~al.}(2011)\citenamefont
  {Barreiro}, \citenamefont {Muller}, \citenamefont {Schindler}, \citenamefont
  {Nigg}, \citenamefont {Monz}, \citenamefont {Chwalla}, \citenamefont
  {Hennrich}, \citenamefont {Roots}, \citenamefont {Zoller},\ and\
  \citenamefont {Blatt}}]{BLA11}%
  \BibitemOpen
  \bibfield  {author} {\bibinfo {author} {\bibfnamefont {J.~T.}\ \bibnamefont
  {Barreiro}}, \bibinfo {author} {\bibfnamefont {M.}~\bibnamefont {Muller}},
  \bibinfo {author} {\bibfnamefont {P.}~\bibnamefont {Schindler}}, \bibinfo
  {author} {\bibfnamefont {D.}~\bibnamefont {Nigg}}, \bibinfo {author}
  {\bibfnamefont {T.}~\bibnamefont {Monz}}, \bibinfo {author} {\bibfnamefont
  {M.}~\bibnamefont {Chwalla}}, \bibinfo {author} {\bibfnamefont
  {M.}~\bibnamefont {Hennrich}}, \bibinfo {author} {\bibfnamefont {C.~F.}\
  \bibnamefont {Roots}}, \bibinfo {author} {\bibfnamefont {P.}~\bibnamefont
  {Zoller}},\ and\ \bibinfo {author} {\bibfnamefont {R.}~\bibnamefont
  {Blatt}},\ }\bibfield  {title} {\bibinfo {title} {An open-system quantum
  simulator with trapped ions},\ }\href@noop {} {\bibfield  {journal} {\bibinfo
   {journal} {Nature}\ }\textbf {\bibinfo {volume} {470}},\ \bibinfo {pages}
  {486} (\bibinfo {year} {2011})}\BibitemShut {NoStop}%
\bibitem [{\citenamefont {Britton}\ \emph {et~al.}(2012)\citenamefont
  {Britton}, \citenamefont {Sawyer}, \citenamefont {Keith}, \citenamefont
  {Wang}, \citenamefont {Freeicks}, \citenamefont {Uys}, \citenamefont
  {Biercuk},\ and\ \citenamefont {Bollinger}}]{BOL12}%
  \BibitemOpen
  \bibfield  {author} {\bibinfo {author} {\bibfnamefont {J.~W.}\ \bibnamefont
  {Britton}}, \bibinfo {author} {\bibfnamefont {B.~C.}\ \bibnamefont {Sawyer}},
  \bibinfo {author} {\bibfnamefont {A.~C.}\ \bibnamefont {Keith}}, \bibinfo
  {author} {\bibfnamefont {C.}~\bibnamefont {Wang}}, \bibinfo {author}
  {\bibfnamefont {J.~K.}\ \bibnamefont {Freeicks}}, \bibinfo {author}
  {\bibfnamefont {H.}~\bibnamefont {Uys}}, \bibinfo {author} {\bibfnamefont
  {M.~J.}\ \bibnamefont {Biercuk}},\ and\ \bibinfo {author} {\bibfnamefont
  {J.~J.}\ \bibnamefont {Bollinger}},\ }\bibfield  {title} {\bibinfo {title}
  {Engineered two-dimensional ising interactions in a trapped-ion quantum
  simulator with hundreds of spins},\ }\href@noop {} {\bibfield  {journal}
  {\bibinfo  {journal} {Nature}\ }\textbf {\bibinfo {volume} {484}},\ \bibinfo
  {pages} {489} (\bibinfo {year} {2012})}\BibitemShut {NoStop}%
\bibitem [{\citenamefont {Baumann}\ \emph {et~al.}(2010)\citenamefont
  {Baumann}, \citenamefont {Guerlin}, \citenamefont {Brennecke},\ and\
  \citenamefont {Esslinger}}]{ESS10}%
  \BibitemOpen
  \bibfield  {author} {\bibinfo {author} {\bibfnamefont {K.}~\bibnamefont
  {Baumann}}, \bibinfo {author} {\bibfnamefont {C.}~\bibnamefont {Guerlin}},
  \bibinfo {author} {\bibfnamefont {F.}~\bibnamefont {Brennecke}},\ and\
  \bibinfo {author} {\bibfnamefont {T.}~\bibnamefont {Esslinger}},\ }\bibfield
  {title} {\bibinfo {title} {Dicke quantum phase transition with a superfluid
  in an optical cavity},\ }\href@noop {} {\bibfield  {journal} {\bibinfo
  {journal} {Nature}\ }\textbf {\bibinfo {volume} {464}},\ \bibinfo {pages}
  {1301} (\bibinfo {year} {2010})}\BibitemShut {NoStop}%
\bibitem [{\citenamefont {Klinder}\ \emph {et~al.}(2015)\citenamefont
  {Klinder}, \citenamefont {Kessler}, \citenamefont {Bakhtiari}, \citenamefont
  {Thorwart},\ and\ \citenamefont {Hemmerich}}]{HEM15}%
  \BibitemOpen
  \bibfield  {author} {\bibinfo {author} {\bibfnamefont {J.}~\bibnamefont
  {Klinder}}, \bibinfo {author} {\bibfnamefont {H.}~\bibnamefont {Kessler}},
  \bibinfo {author} {\bibfnamefont {M.~R.}\ \bibnamefont {Bakhtiari}}, \bibinfo
  {author} {\bibfnamefont {M.}~\bibnamefont {Thorwart}},\ and\ \bibinfo
  {author} {\bibfnamefont {A.}~\bibnamefont {Hemmerich}},\ }\bibfield  {title}
  {\bibinfo {title} {Observation of a superradiant mott insulator in the
  dicke-hubbard model},\ }\href@noop {} {\bibfield  {journal} {\bibinfo
  {journal} {Phys. Rev. Lett.}\ }\textbf {\bibinfo {volume} {115}},\ \bibinfo
  {pages} {230403} (\bibinfo {year} {2015})}\BibitemShut {NoStop}%
\bibitem [{\citenamefont {Leonard}\ \emph {et~al.}(2017)\citenamefont
  {Leonard}, \citenamefont {Morales}, \citenamefont {Zupancic}, \citenamefont
  {Esslinger},\ and\ \citenamefont {Donner}}]{DON17}%
  \BibitemOpen
  \bibfield  {author} {\bibinfo {author} {\bibfnamefont {J.}~\bibnamefont
  {Leonard}}, \bibinfo {author} {\bibfnamefont {A.}~\bibnamefont {Morales}},
  \bibinfo {author} {\bibfnamefont {P.}~\bibnamefont {Zupancic}}, \bibinfo
  {author} {\bibfnamefont {T.}~\bibnamefont {Esslinger}},\ and\ \bibinfo
  {author} {\bibfnamefont {T.}~\bibnamefont {Donner}},\ }\bibfield  {title}
  {\bibinfo {title} {Supersolid formation in a quantum gas breaking a
  continuous translation symmetry},\ }\href@noop {} {\bibfield  {journal}
  {\bibinfo  {journal} {Nature}\ }\textbf {\bibinfo {volume} {543}},\ \bibinfo
  {pages} {87} (\bibinfo {year} {2017})}\BibitemShut {NoStop}%
\bibitem [{\citenamefont {Norcia}\ \emph {et~al.}(2018)\citenamefont {Norcia},
  \citenamefont {Lewis-Swan}, \citenamefont {Cline}, \citenamefont {Zhu},
  \citenamefont {Rey},\ and\ \citenamefont {Thompson}}]{THO18}%
  \BibitemOpen
  \bibfield  {author} {\bibinfo {author} {\bibfnamefont {M.~A.}\ \bibnamefont
  {Norcia}}, \bibinfo {author} {\bibfnamefont {R.~J.}\ \bibnamefont
  {Lewis-Swan}}, \bibinfo {author} {\bibfnamefont {J.}~\bibnamefont {Cline}},
  \bibinfo {author} {\bibfnamefont {B.}~\bibnamefont {Zhu}}, \bibinfo {author}
  {\bibfnamefont {A.~M.}\ \bibnamefont {Rey}},\ and\ \bibinfo {author}
  {\bibfnamefont {J.~K.}\ \bibnamefont {Thompson}},\ }\bibfield  {title}
  {\bibinfo {title} {Cavity-mediated collective spin-exchange interactions in a
  strontium superradiant laser},\ }\href@noop {} {\bibfield  {journal}
  {\bibinfo  {journal} {Science}\ }\textbf {\bibinfo {volume} {361}},\ \bibinfo
  {pages} {6399} (\bibinfo {year} {2018})}\BibitemShut {NoStop}%
\bibitem [{\citenamefont {Vaidya}\ \emph {et~al.}(2018)\citenamefont {Vaidya},
  \citenamefont {Guo}, \citenamefont {Kroeze}, \citenamefont {Ballantine},
  \citenamefont {Kollar}, \citenamefont {Keeling},\ and\ \citenamefont
  {Lev}}]{LEV18}%
  \BibitemOpen
  \bibfield  {author} {\bibinfo {author} {\bibfnamefont {V.~D.}\ \bibnamefont
  {Vaidya}}, \bibinfo {author} {\bibfnamefont {Y.}~\bibnamefont {Guo}},
  \bibinfo {author} {\bibfnamefont {R.~M.}\ \bibnamefont {Kroeze}}, \bibinfo
  {author} {\bibfnamefont {K.~E.}\ \bibnamefont {Ballantine}}, \bibinfo
  {author} {\bibfnamefont {A.~J.}\ \bibnamefont {Kollar}}, \bibinfo {author}
  {\bibfnamefont {J.}~\bibnamefont {Keeling}},\ and\ \bibinfo {author}
  {\bibfnamefont {B.~L.}\ \bibnamefont {Lev}},\ }\bibfield  {title} {\bibinfo
  {title} {Tunable-range, photon-mediated atomic interactions in multimode
  cavity qed},\ }\href@noop {} {\bibfield  {journal} {\bibinfo  {journal}
  {Phys. Rev. X}\ }\textbf {\bibinfo {volume} {8}},\ \bibinfo {pages} {011002}
  (\bibinfo {year} {2018})}\BibitemShut {NoStop}%
\bibitem [{\citenamefont {Guo}\ \emph {et~al.}(2019)\citenamefont {Guo},
  \citenamefont {Kroeze}, \citenamefont {Vaidya}, \citenamefont {Keeling},\
  and\ \citenamefont {Lev}}]{LEV19}%
  \BibitemOpen
  \bibfield  {author} {\bibinfo {author} {\bibfnamefont {Y.}~\bibnamefont
  {Guo}}, \bibinfo {author} {\bibfnamefont {R.~M.}\ \bibnamefont {Kroeze}},
  \bibinfo {author} {\bibfnamefont {V.~D.}\ \bibnamefont {Vaidya}}, \bibinfo
  {author} {\bibfnamefont {J.}~\bibnamefont {Keeling}},\ and\ \bibinfo {author}
  {\bibfnamefont {B.~L.}\ \bibnamefont {Lev}},\ }\bibfield  {title} {\bibinfo
  {title} {Sign-changing phonon-mediated atom interactions in multimode cavity
  quantum electrodynamics},\ }\href@noop {} {\bibfield  {journal} {\bibinfo
  {journal} {Phys. Rev. Lett.}\ }\textbf {\bibinfo {volume} {122}},\ \bibinfo
  {pages} {193601} (\bibinfo {year} {2019})}\BibitemShut {NoStop}%
\bibitem [{\citenamefont {DeSalvo}\ \emph {et~al.}(2019)\citenamefont
  {DeSalvo}, \citenamefont {Patel}, \citenamefont {Cai},\ and\ \citenamefont
  {Chin}}]{CHI19}%
  \BibitemOpen
  \bibfield  {author} {\bibinfo {author} {\bibfnamefont {B.~J.}\ \bibnamefont
  {DeSalvo}}, \bibinfo {author} {\bibfnamefont {K.}~\bibnamefont {Patel}},
  \bibinfo {author} {\bibfnamefont {G.}~\bibnamefont {Cai}},\ and\ \bibinfo
  {author} {\bibfnamefont {C.}~\bibnamefont {Chin}},\ }\bibfield  {title}
  {\bibinfo {title} {Observation of fermion-mediated interactions between
  bosonic atoms},\ }\href@noop {} {\bibfield  {journal} {\bibinfo  {journal}
  {Nature}\ }\textbf {\bibinfo {volume} {568}},\ \bibinfo {pages} {61}
  (\bibinfo {year} {2019})}\BibitemShut {NoStop}%
\bibitem [{\citenamefont {Bruun}(2019)}]{BRU19}%
  \BibitemOpen
  \bibfield  {author} {\bibinfo {author} {\bibfnamefont {G.~M.}\ \bibnamefont
  {Bruun}},\ }\bibfield  {title} {\bibinfo {title} {New interactions seen in an
  ultracold gas},\ }\href@noop {} {\bibfield  {journal} {\bibinfo  {journal}
  {Nature}\ }\textbf {\bibinfo {volume} {568}},\ \bibinfo {pages} {37}
  (\bibinfo {year} {2019})}\BibitemShut {NoStop}%
\bibitem [{\citenamefont {Edri}\ \emph {et~al.}(2020)\citenamefont {Edri},
  \citenamefont {Raz}, \citenamefont {Matzliah}, \citenamefont {Davidson},\
  and\ \citenamefont {Ozeri}}]{OZE20}%
  \BibitemOpen
  \bibfield  {author} {\bibinfo {author} {\bibfnamefont {H.}~\bibnamefont
  {Edri}}, \bibinfo {author} {\bibfnamefont {B.}~\bibnamefont {Raz}}, \bibinfo
  {author} {\bibfnamefont {N.}~\bibnamefont {Matzliah}}, \bibinfo {author}
  {\bibfnamefont {N.}~\bibnamefont {Davidson}},\ and\ \bibinfo {author}
  {\bibfnamefont {R.}~\bibnamefont {Ozeri}},\ }\bibfield  {title} {\bibinfo
  {title} {Observation of spin-spin fermion-mediated interactions between
  ultracold atoms},\ }\href@noop {} {\bibfield  {journal} {\bibinfo  {journal}
  {Phys. Rev. Lett.}\ }\textbf {\bibinfo {volume} {124}},\ \bibinfo {pages}
  {163401} (\bibinfo {year} {2020})}\BibitemShut {NoStop}%
\bibitem [{\citenamefont {Santamore}\ and\ \citenamefont
  {Timmermans}(2008)}]{TIM08}%
  \BibitemOpen
  \bibfield  {author} {\bibinfo {author} {\bibfnamefont {D.~H.}\ \bibnamefont
  {Santamore}}\ and\ \bibinfo {author} {\bibfnamefont {E.}~\bibnamefont
  {Timmermans}},\ }\bibfield  {title} {\bibinfo {title} {Fermion-mediated
  interations in a dilute bose-einstein condenstate},\ }\href@noop {}
  {\bibfield  {journal} {\bibinfo  {journal} {Phys. Rev. A}\ }\textbf {\bibinfo
  {volume} {78}},\ \bibinfo {pages} {013619} (\bibinfo {year}
  {2008})}\BibitemShut {NoStop}%
\bibitem [{\citenamefont {De}\ and\ \citenamefont {B.}(2014)}]{SPI14}%
  \BibitemOpen
  \bibfield  {author} {\bibinfo {author} {\bibfnamefont {S.}~\bibnamefont
  {De}}\ and\ \bibinfo {author} {\bibfnamefont {S.~I.}\ \bibnamefont {B.}},\
  }\bibfield  {title} {\bibinfo {title} {Fermion-mediated long-range
  interactions between bosons stored in an optical lattice},\ }\href@noop {}
  {\bibfield  {journal} {\bibinfo  {journal} {Appl. Phys. B}\ }\textbf
  {\bibinfo {volume} {114}},\ \bibinfo {pages} {527} (\bibinfo {year}
  {2014})}\BibitemShut {NoStop}%
\bibitem [{\citenamefont {Kinnunen}\ and\ \citenamefont {Bruun}(2015)}]{BRU15}%
  \BibitemOpen
  \bibfield  {author} {\bibinfo {author} {\bibfnamefont {J.~J.}\ \bibnamefont
  {Kinnunen}}\ and\ \bibinfo {author} {\bibfnamefont {G.~M.}\ \bibnamefont
  {Bruun}},\ }\bibfield  {title} {\bibinfo {title} {Induced interactions in a
  superfluid bose-fermi mixture},\ }\href@noop {} {\bibfield  {journal}
  {\bibinfo  {journal} {Phys. Rev. A}\ }\textbf {\bibinfo {volume} {91}},\
  \bibinfo {pages} {041605(R)} (\bibinfo {year} {2015})}\BibitemShut {NoStop}%
\bibitem [{\citenamefont {Ho}\ and\ \citenamefont {Shenoy}(1996)}]{SHE96}%
  \BibitemOpen
  \bibfield  {author} {\bibinfo {author} {\bibfnamefont {T.-L.}\ \bibnamefont
  {Ho}}\ and\ \bibinfo {author} {\bibfnamefont {V.~B.}\ \bibnamefont
  {Shenoy}},\ }\bibfield  {title} {\bibinfo {title} {Binary mixtures of bose
  condensates of alkali atoms},\ }\href@noop {} {\bibfield  {journal} {\bibinfo
   {journal} {Phys. Rev. Lett.}\ }\textbf {\bibinfo {volume} {77}},\ \bibinfo
  {pages} {3276} (\bibinfo {year} {1996})}\BibitemShut {NoStop}%
\bibitem [{\citenamefont {Pu}\ and\ \citenamefont {Bigelow}(1998)}]{BIG98}%
  \BibitemOpen
  \bibfield  {author} {\bibinfo {author} {\bibfnamefont {H.}~\bibnamefont
  {Pu}}\ and\ \bibinfo {author} {\bibfnamefont {N.~P.}\ \bibnamefont
  {Bigelow}},\ }\bibfield  {title} {\bibinfo {title} {Properties of two-species
  bose condensates},\ }\href@noop {} {\bibfield  {journal} {\bibinfo  {journal}
  {Phys. Rev. Lett.}\ }\textbf {\bibinfo {volume} {80}},\ \bibinfo {pages}
  {1130} (\bibinfo {year} {1998})}\BibitemShut {NoStop}%
\bibitem [{\citenamefont {Ao}\ and\ \citenamefont {Chui}(1998)}]{CHU98}%
  \BibitemOpen
  \bibfield  {author} {\bibinfo {author} {\bibfnamefont {P.}~\bibnamefont
  {Ao}}\ and\ \bibinfo {author} {\bibfnamefont {S.~T.}\ \bibnamefont {Chui}},\
  }\bibfield  {title} {\bibinfo {title} {Binary bose-einstein condensate
  mixtures in weakly and strongly segregated phases},\ }\href@noop {}
  {\bibfield  {journal} {\bibinfo  {journal} {Phys. Rev. A}\ }\textbf {\bibinfo
  {volume} {58}},\ \bibinfo {pages} {4836} (\bibinfo {year}
  {1998})}\BibitemShut {NoStop}%
\bibitem [{\citenamefont {Timmermans}(1998)}]{TIM98}%
  \BibitemOpen
  \bibfield  {author} {\bibinfo {author} {\bibfnamefont {E.}~\bibnamefont
  {Timmermans}},\ }\bibfield  {title} {\bibinfo {title} {Phase separation of
  bose-einstein condensates},\ }\href@noop {} {\bibfield  {journal} {\bibinfo
  {journal} {Phys. Rev. Lett.}\ }\textbf {\bibinfo {volume} {81}},\ \bibinfo
  {pages} {5718} (\bibinfo {year} {1998})}\BibitemShut {NoStop}%
\bibitem [{\citenamefont {Chien}\ \emph {et~al.}(2012)\citenamefont {Chien},
  \citenamefont {Cooper},\ and\ \citenamefont {Timmermans}}]{TIM12}%
  \BibitemOpen
  \bibfield  {author} {\bibinfo {author} {\bibfnamefont {C.-C.}\ \bibnamefont
  {Chien}}, \bibinfo {author} {\bibfnamefont {F.}~\bibnamefont {Cooper}},\ and\
  \bibinfo {author} {\bibfnamefont {E.}~\bibnamefont {Timmermans}},\ }\bibfield
   {title} {\bibinfo {title} {Large-n approximation for one- and two-component
  dilute bose gases},\ }\href@noop {} {\bibfield  {journal} {\bibinfo
  {journal} {Phys. Rev. A}\ }\textbf {\bibinfo {volume} {86}},\ \bibinfo
  {pages} {023634} (\bibinfo {year} {2012})}\BibitemShut {NoStop}%
\bibitem [{\citenamefont {Armaitis}\ \emph {et~al.}(2015)\citenamefont
  {Armaitis}, \citenamefont {Stoof},\ and\ \citenamefont {Duine}}]{DUI15}%
  \BibitemOpen
  \bibfield  {author} {\bibinfo {author} {\bibfnamefont {J.}~\bibnamefont
  {Armaitis}}, \bibinfo {author} {\bibfnamefont {H.~T.~C.}\ \bibnamefont
  {Stoof}},\ and\ \bibinfo {author} {\bibfnamefont {R.~A.}\ \bibnamefont
  {Duine}},\ }\bibfield  {title} {\bibinfo {title} {Hydrodynamic modes of
  partially condensed bose mixtures},\ }\href@noop {} {\bibfield  {journal}
  {\bibinfo  {journal} {Phys. Rev. A}\ }\textbf {\bibinfo {volume} {91}},\
  \bibinfo {pages} {043641} (\bibinfo {year} {2015})}\BibitemShut {NoStop}%
\bibitem [{\citenamefont {Lee}\ \emph {et~al.}(2016)\citenamefont {Lee},
  \citenamefont {Jorgensen}, \citenamefont {Liu}, \citenamefont {Wacker},
  \citenamefont {Arlt},\ and\ \citenamefont {Proukakis}}]{PRO16}%
  \BibitemOpen
  \bibfield  {author} {\bibinfo {author} {\bibfnamefont {K.~L.}\ \bibnamefont
  {Lee}}, \bibinfo {author} {\bibfnamefont {N.~B.}\ \bibnamefont {Jorgensen}},
  \bibinfo {author} {\bibfnamefont {I.~K.}\ \bibnamefont {Liu}}, \bibinfo
  {author} {\bibfnamefont {L.}~\bibnamefont {Wacker}}, \bibinfo {author}
  {\bibfnamefont {J.~J.}\ \bibnamefont {Arlt}},\ and\ \bibinfo {author}
  {\bibfnamefont {N.~P.}\ \bibnamefont {Proukakis}},\ }\bibfield  {title}
  {\bibinfo {title} {Phase separation and dynamics of two-component
  bose-einstein condensates},\ }\href@noop {} {\bibfield  {journal} {\bibinfo
  {journal} {Phys. Rev. A}\ }\textbf {\bibinfo {volume} {94}},\ \bibinfo
  {pages} {013602} (\bibinfo {year} {2016})}\BibitemShut {NoStop}%
\bibitem [{\citenamefont {Ota}\ \emph {et~al.}(2019)\citenamefont {Ota},
  \citenamefont {Giorgini},\ and\ \citenamefont {Stringari}}]{STR19}%
  \BibitemOpen
  \bibfield  {author} {\bibinfo {author} {\bibfnamefont {M.}~\bibnamefont
  {Ota}}, \bibinfo {author} {\bibfnamefont {S.}~\bibnamefont {Giorgini}},\ and\
  \bibinfo {author} {\bibfnamefont {S.}~\bibnamefont {Stringari}},\ }\bibfield
  {title} {\bibinfo {title} {Magnetic phase transition in a mixture of two
  interesting superfluid bose gases at finite temperature},\ }\href@noop {}
  {\bibfield  {journal} {\bibinfo  {journal} {Phys. Rev. Lett.}\ }\textbf
  {\bibinfo {volume} {123}},\ \bibinfo {pages} {075301} (\bibinfo {year}
  {2019})}\BibitemShut {NoStop}%
\bibitem [{\citenamefont {Wen}\ \emph {et~al.}(2020)\citenamefont {Wen},
  \citenamefont {Guo}, \citenamefont {Wang}, \citenamefont {Hu}, \citenamefont
  {Saito}, \citenamefont {Dai},\ and\ \citenamefont {Zhang}}]{WEN20}%
  \BibitemOpen
  \bibfield  {author} {\bibinfo {author} {\bibfnamefont {L.}~\bibnamefont
  {Wen}}, \bibinfo {author} {\bibfnamefont {H.}~\bibnamefont {Guo}}, \bibinfo
  {author} {\bibfnamefont {Y.-J.}\ \bibnamefont {Wang}}, \bibinfo {author}
  {\bibfnamefont {A.-Y.}\ \bibnamefont {Hu}}, \bibinfo {author} {\bibfnamefont
  {H.}~\bibnamefont {Saito}}, \bibinfo {author} {\bibfnamefont {C.-Q.}\
  \bibnamefont {Dai}},\ and\ \bibinfo {author} {\bibfnamefont {X.-F.}\
  \bibnamefont {Zhang}},\ }\bibfield  {title} {\bibinfo {title} {Effects of
  atom numbers on the miscibility-immiscibility transition of a binary
  bose-einstein condensate},\ }\href@noop {} {\bibfield  {journal} {\bibinfo
  {journal} {Phys. Rev. A}\ }\textbf {\bibinfo {volume} {101}},\ \bibinfo
  {pages} {033610} (\bibinfo {year} {2020})}\BibitemShut {NoStop}%
\bibitem [{\citenamefont {Hall}\ \emph {et~al.}(1998)\citenamefont {Hall},
  \citenamefont {Matthews}, \citenamefont {Ensher}, \citenamefont {Wieman},\
  and\ \citenamefont {Cornell}}]{COR98}%
  \BibitemOpen
  \bibfield  {author} {\bibinfo {author} {\bibfnamefont {D.~S.}\ \bibnamefont
  {Hall}}, \bibinfo {author} {\bibfnamefont {M.~R.}\ \bibnamefont {Matthews}},
  \bibinfo {author} {\bibfnamefont {J.~R.}\ \bibnamefont {Ensher}}, \bibinfo
  {author} {\bibfnamefont {C.~E.}\ \bibnamefont {Wieman}},\ and\ \bibinfo
  {author} {\bibfnamefont {E.~A.}\ \bibnamefont {Cornell}},\ }\bibfield
  {title} {\bibinfo {title} {Dynamics of component separation in binary mixture
  of bose-einstein condensates},\ }\href@noop {} {\bibfield  {journal}
  {\bibinfo  {journal} {Phys. Rev. Lett.}\ }\textbf {\bibinfo {volume} {81}},\
  \bibinfo {pages} {1539} (\bibinfo {year} {1998})}\BibitemShut {NoStop}%
\bibitem [{\citenamefont {Papp}\ \emph {et~al.}(2008)\citenamefont {Papp},
  \citenamefont {Pino},\ and\ \citenamefont {Wieman}}]{WIE08}%
  \BibitemOpen
  \bibfield  {author} {\bibinfo {author} {\bibfnamefont {S.~B.}\ \bibnamefont
  {Papp}}, \bibinfo {author} {\bibfnamefont {J.~M.}\ \bibnamefont {Pino}},\
  and\ \bibinfo {author} {\bibfnamefont {C.~E.}\ \bibnamefont {Wieman}},\
  }\bibfield  {title} {\bibinfo {title} {Tunable miscibility in a dual-species
  bose-einstein condensate},\ }\href@noop {} {\bibfield  {journal} {\bibinfo
  {journal} {Phys. Rev. Lett.}\ }\textbf {\bibinfo {volume} {101}},\ \bibinfo
  {pages} {040402} (\bibinfo {year} {2008})}\BibitemShut {NoStop}%
\bibitem [{\citenamefont {Tojo}\ \emph {et~al.}(2010)\citenamefont {Tojo},
  \citenamefont {Taguchi}, \citenamefont {Masuyama}, \citenamefont {Hayashi},
  \citenamefont {Saito},\ and\ \citenamefont {Hirano}}]{HIR10}%
  \BibitemOpen
  \bibfield  {author} {\bibinfo {author} {\bibfnamefont {S.}~\bibnamefont
  {Tojo}}, \bibinfo {author} {\bibfnamefont {Y.}~\bibnamefont {Taguchi}},
  \bibinfo {author} {\bibfnamefont {Y.}~\bibnamefont {Masuyama}}, \bibinfo
  {author} {\bibfnamefont {T.}~\bibnamefont {Hayashi}}, \bibinfo {author}
  {\bibfnamefont {H.}~\bibnamefont {Saito}},\ and\ \bibinfo {author}
  {\bibfnamefont {T.}~\bibnamefont {Hirano}},\ }\bibfield  {title} {\bibinfo
  {title} {Controlling phase separation of binary bose-einstein condensates via
  mixed-spin channel feshbach resonance},\ }\href@noop {} {\bibfield  {journal}
  {\bibinfo  {journal} {Phys. Rev. A}\ }\textbf {\bibinfo {volume} {82}},\
  \bibinfo {pages} {033609} (\bibinfo {year} {2010})}\BibitemShut {NoStop}%
\bibitem [{\citenamefont {Nicklas}\ \emph {et~al.}(2011)\citenamefont
  {Nicklas}, \citenamefont {Strobel}, \citenamefont {Zibold}, \citenamefont
  {Gross}, \citenamefont {Malomed}, \citenamefont {Kevrekidis},\ and\
  \citenamefont {Oberthaler}}]{OBE11}%
  \BibitemOpen
  \bibfield  {author} {\bibinfo {author} {\bibfnamefont {E.}~\bibnamefont
  {Nicklas}}, \bibinfo {author} {\bibfnamefont {H.}~\bibnamefont {Strobel}},
  \bibinfo {author} {\bibfnamefont {T.}~\bibnamefont {Zibold}}, \bibinfo
  {author} {\bibfnamefont {C.}~\bibnamefont {Gross}}, \bibinfo {author}
  {\bibfnamefont {B.~A.}\ \bibnamefont {Malomed}}, \bibinfo {author}
  {\bibfnamefont {P.~G.}\ \bibnamefont {Kevrekidis}},\ and\ \bibinfo {author}
  {\bibfnamefont {M.~K.}\ \bibnamefont {Oberthaler}},\ }\bibfield  {title}
  {\bibinfo {title} {Rabi flopping induces spatial demixing dynamics},\
  }\href@noop {} {\bibfield  {journal} {\bibinfo  {journal} {Phys. Rev. Lett.}\
  }\textbf {\bibinfo {volume} {107}},\ \bibinfo {pages} {193001} (\bibinfo
  {year} {2011})}\BibitemShut {NoStop}%
\bibitem [{\citenamefont {Pasquiou}\ \emph {et~al.}(2013)\citenamefont
  {Pasquiou}, \citenamefont {Bayerle}, \citenamefont {Tzanova}, \citenamefont
  {Stellmer}, \citenamefont {Szczepkowski}, \citenamefont {Parigger},
  \citenamefont {Grimm},\ and\ \citenamefont {Schreck}}]{SCH13}%
  \BibitemOpen
  \bibfield  {author} {\bibinfo {author} {\bibfnamefont {B.}~\bibnamefont
  {Pasquiou}}, \bibinfo {author} {\bibfnamefont {A.}~\bibnamefont {Bayerle}},
  \bibinfo {author} {\bibfnamefont {S.~M.}\ \bibnamefont {Tzanova}}, \bibinfo
  {author} {\bibfnamefont {S.}~\bibnamefont {Stellmer}}, \bibinfo {author}
  {\bibfnamefont {J.}~\bibnamefont {Szczepkowski}}, \bibinfo {author}
  {\bibfnamefont {M.}~\bibnamefont {Parigger}}, \bibinfo {author}
  {\bibfnamefont {R.}~\bibnamefont {Grimm}},\ and\ \bibinfo {author}
  {\bibfnamefont {F.}~\bibnamefont {Schreck}},\ }\bibfield  {title} {\bibinfo
  {title} {Quantum degenerate mixtures of strontium and rubidium atoms},\
  }\href@noop {} {\bibfield  {journal} {\bibinfo  {journal} {Phys. Rev. A}\
  }\textbf {\bibinfo {volume} {88}},\ \bibinfo {pages} {023601} (\bibinfo
  {year} {2013})}\BibitemShut {NoStop}%
\bibitem [{\citenamefont {Fava}\ \emph {et~al.}(2018)\citenamefont {Fava},
  \citenamefont {Bienaime}, \citenamefont {Mordini}, \citenamefont {Colzi},
  \citenamefont {Qu}, \citenamefont {Stringari}, \citenamefont {Lamporesi},\
  and\ \citenamefont {Ferrari}}]{FER18}%
  \BibitemOpen
  \bibfield  {author} {\bibinfo {author} {\bibfnamefont {E.}~\bibnamefont
  {Fava}}, \bibinfo {author} {\bibfnamefont {T.}~\bibnamefont {Bienaime}},
  \bibinfo {author} {\bibfnamefont {C.}~\bibnamefont {Mordini}}, \bibinfo
  {author} {\bibfnamefont {G.}~\bibnamefont {Colzi}}, \bibinfo {author}
  {\bibfnamefont {C.}~\bibnamefont {Qu}}, \bibinfo {author} {\bibfnamefont
  {S.}~\bibnamefont {Stringari}}, \bibinfo {author} {\bibfnamefont
  {G.}~\bibnamefont {Lamporesi}},\ and\ \bibinfo {author} {\bibfnamefont
  {G.}~\bibnamefont {Ferrari}},\ }\bibfield  {title} {\bibinfo {title}
  {Observation of spin superfluidity in a bose gas mixture},\ }\href@noop {}
  {\bibfield  {journal} {\bibinfo  {journal} {Phys. Rev. Lett.}\ }\textbf
  {\bibinfo {volume} {120}},\ \bibinfo {pages} {170401} (\bibinfo {year}
  {2018})}\BibitemShut {NoStop}%
\bibitem [{\citenamefont {Lee}\ \emph {et~al.}(2018)\citenamefont {Lee},
  \citenamefont {Jorgensen}, \citenamefont {Wacker}, \citenamefont {Skou},
  \citenamefont {Skalmstang}, \citenamefont {Arlt},\ and\ \citenamefont
  {Proukakis}}]{PRO18}%
  \BibitemOpen
  \bibfield  {author} {\bibinfo {author} {\bibfnamefont {K.~L.}\ \bibnamefont
  {Lee}}, \bibinfo {author} {\bibfnamefont {N.~B.}\ \bibnamefont {Jorgensen}},
  \bibinfo {author} {\bibfnamefont {L.~J.}\ \bibnamefont {Wacker}}, \bibinfo
  {author} {\bibfnamefont {M.~G.}\ \bibnamefont {Skou}}, \bibinfo {author}
  {\bibfnamefont {K.~T.}\ \bibnamefont {Skalmstang}}, \bibinfo {author}
  {\bibfnamefont {J.~J.}\ \bibnamefont {Arlt}},\ and\ \bibinfo {author}
  {\bibfnamefont {N.~P.}\ \bibnamefont {Proukakis}},\ }\bibfield  {title}
  {\bibinfo {title} {Time-of-flight expansion of binary bose-einstein
  condensates at finite temperature},\ }\href@noop {} {\bibfield  {journal}
  {\bibinfo  {journal} {New J. Phys.}\ }\textbf {\bibinfo {volume} {20}},\
  \bibinfo {pages} {053004} (\bibinfo {year} {2018})}\BibitemShut {NoStop}%
\bibitem [{\citenamefont {Mukherjee}\ \emph {et~al.}(2017)\citenamefont
  {Mukherjee}, \citenamefont {Yan}, \citenamefont {Patel}, \citenamefont
  {Hadzibabic}, \citenamefont {Yefsah}, \citenamefont {Struck},\ and\
  \citenamefont {Zwierlein}}]{ZWI17}%
  \BibitemOpen
  \bibfield  {author} {\bibinfo {author} {\bibfnamefont {B.}~\bibnamefont
  {Mukherjee}}, \bibinfo {author} {\bibfnamefont {Z.}~\bibnamefont {Yan}},
  \bibinfo {author} {\bibfnamefont {P.~B.}\ \bibnamefont {Patel}}, \bibinfo
  {author} {\bibfnamefont {Z.}~\bibnamefont {Hadzibabic}}, \bibinfo {author}
  {\bibfnamefont {T.}~\bibnamefont {Yefsah}}, \bibinfo {author} {\bibfnamefont
  {J.}~\bibnamefont {Struck}},\ and\ \bibinfo {author} {\bibfnamefont {M.~W.}\
  \bibnamefont {Zwierlein}},\ }\bibfield  {title} {\bibinfo {title}
  {Homogeneous atomic fermi gases},\ }\href@noop {} {\bibfield  {journal}
  {\bibinfo  {journal} {Phys. Rev. Lett.}\ }\textbf {\bibinfo {volume} {118}},\
  \bibinfo {pages} {123401} (\bibinfo {year} {2017})}\BibitemShut {NoStop}%
\bibitem [{\citenamefont {Lopes}\ \emph {et~al.}(2017)\citenamefont {Lopes},
  \citenamefont {Eigen}, \citenamefont {Navon}, \citenamefont {Clement},
  \citenamefont {Smith},\ and\ \citenamefont {Hadzibabic}}]{HAD17}%
  \BibitemOpen
  \bibfield  {author} {\bibinfo {author} {\bibfnamefont {R.}~\bibnamefont
  {Lopes}}, \bibinfo {author} {\bibfnamefont {C.}~\bibnamefont {Eigen}},
  \bibinfo {author} {\bibfnamefont {N.}~\bibnamefont {Navon}}, \bibinfo
  {author} {\bibfnamefont {D.}~\bibnamefont {Clement}}, \bibinfo {author}
  {\bibfnamefont {R.~P.}\ \bibnamefont {Smith}},\ and\ \bibinfo {author}
  {\bibfnamefont {Z.}~\bibnamefont {Hadzibabic}},\ }\bibfield  {title}
  {\bibinfo {title} {Quantum deplection of a homogeneous bose-einstein
  condensate},\ }\href@noop {} {\bibfield  {journal} {\bibinfo  {journal}
  {Phys. Rev. Lett.}\ }\textbf {\bibinfo {volume} {119}},\ \bibinfo {pages}
  {190404} (\bibinfo {year} {2017})}\BibitemShut {NoStop}%
\bibitem [{\citenamefont {Altland}\ and\ \citenamefont {Simons}(2010)}]{SIM10}%
  \BibitemOpen
  \bibfield  {author} {\bibinfo {author} {\bibfnamefont {A.}~\bibnamefont
  {Altland}}\ and\ \bibinfo {author} {\bibfnamefont {B.}~\bibnamefont
  {Simons}},\ }\href@noop {} {\emph {\bibinfo {title} {Condensed Matter Field
  Theory}}},\ \bibinfo {edition} {2nd}\ ed.\ (\bibinfo  {publisher} {CUP},\
  \bibinfo {address} {Cambridge, UK},\ \bibinfo {year} {2010})\BibitemShut
  {NoStop}%
\bibitem [{\citenamefont {Hugenholz}\ and\ \citenamefont {Pines}(1959)}]{HP59}%
  \BibitemOpen
  \bibfield  {author} {\bibinfo {author} {\bibfnamefont {N.}~\bibnamefont
  {Hugenholz}}\ and\ \bibinfo {author} {\bibfnamefont {D.}~\bibnamefont
  {Pines}},\ }\bibfield  {title} {\bibinfo {title} {Ground-state energy and
  excitation spectrum of a system of interacting bosons},\ }\href
  {https://doi.org/10.1103/PhysRev.116.489} {\bibfield  {journal} {\bibinfo
  {journal} {Phys. Rev.}\ }\textbf {\bibinfo {volume} {116}},\ \bibinfo {pages}
  {489} (\bibinfo {year} {1959})}\BibitemShut {NoStop}%
\bibitem [{\citenamefont {Ruderman}\ and\ \citenamefont
  {Kittel}(1954)}]{RKKY54}%
  \BibitemOpen
  \bibfield  {author} {\bibinfo {author} {\bibfnamefont {M.~A.}\ \bibnamefont
  {Ruderman}}\ and\ \bibinfo {author} {\bibfnamefont {C.}~\bibnamefont
  {Kittel}},\ }\bibfield  {title} {\bibinfo {title} {Indirect exchange coupling
  of nuclear magnetic moments by conduction electrons},\ }\href@noop {}
  {\bibfield  {journal} {\bibinfo  {journal} {Phys. Rev.}\ }\textbf {\bibinfo
  {volume} {96}},\ \bibinfo {pages} {99} (\bibinfo {year} {1954})}\BibitemShut
  {NoStop}%
\bibitem [{\citenamefont {Lindhard}(1954)}]{LIN54}%
  \BibitemOpen
  \bibfield  {author} {\bibinfo {author} {\bibfnamefont {J.}~\bibnamefont
  {Lindhard}},\ }\bibfield  {title} {\bibinfo {title} {On the properties of a
  gas charged particles},\ }\href@noop {} {\bibfield  {journal} {\bibinfo
  {journal} {K. Dan. Vidensk. Selsk. Mat. Fys Medd.}\ }\textbf {\bibinfo
  {volume} {28}},\ \bibinfo {pages} {8} (\bibinfo {year} {1954})}\BibitemShut
  {NoStop}%
\bibitem [{\citenamefont {M{\o}lmer}(1998)}]{MOL98}%
  \BibitemOpen
  \bibfield  {author} {\bibinfo {author} {\bibfnamefont {K.}~\bibnamefont
  {M{\o}lmer}},\ }\bibfield  {title} {\bibinfo {title} {Bose condensates and
  fermi gases at zero temperature},\ }\href@noop {} {\bibfield  {journal}
  {\bibinfo  {journal} {Phys. Rev. Lett.}\ }\textbf {\bibinfo {volume} {80}},\
  \bibinfo {pages} {1804} (\bibinfo {year} {1998})}\BibitemShut {NoStop}%
\bibitem [{\citenamefont {Viverit}\ \emph {et~al.}(2000)\citenamefont
  {Viverit}, \citenamefont {Pethick},\ and\ \citenamefont {Smith}}]{VIV00}%
  \BibitemOpen
  \bibfield  {author} {\bibinfo {author} {\bibfnamefont {L.}~\bibnamefont
  {Viverit}}, \bibinfo {author} {\bibfnamefont {C.~J.}\ \bibnamefont
  {Pethick}},\ and\ \bibinfo {author} {\bibfnamefont {H.}~\bibnamefont
  {Smith}},\ }\bibfield  {title} {\bibinfo {title} {Zero-temperature phase
  diagram of binary boson-fermion mixtures},\ }\href@noop {} {\bibfield
  {journal} {\bibinfo  {journal} {Phys. Rev. A}\ }\textbf {\bibinfo {volume}
  {61}},\ \bibinfo {pages} {053605} (\bibinfo {year} {2000})}\BibitemShut
  {NoStop}%
\bibitem [{\citenamefont {Roth}(2002)}]{ROT02}%
  \BibitemOpen
  \bibfield  {author} {\bibinfo {author} {\bibfnamefont {R.}~\bibnamefont
  {Roth}},\ }\bibfield  {title} {\bibinfo {title} {Struture and stability of
  trapped atomic boson-fermion mixture},\ }\href@noop {} {\bibfield  {journal}
  {\bibinfo  {journal} {Phys. Rev. A}\ }\textbf {\bibinfo {volume} {66}},\
  \bibinfo {pages} {013614} (\bibinfo {year} {2002})}\BibitemShut {NoStop}%
\bibitem [{\citenamefont {Liu}(1997)}]{LIU97}%
  \BibitemOpen
  \bibfield  {author} {\bibinfo {author} {\bibfnamefont {W.~V.}\ \bibnamefont
  {Liu}},\ }\bibfield  {title} {\bibinfo {title} {Theoretical study of the
  damping of collective excitations in a bose-einstein condensate},\
  }\href@noop {} {\bibfield  {journal} {\bibinfo  {journal} {Phys. Rev. Lett.}\
  }\textbf {\bibinfo {volume} {79}},\ \bibinfo {pages} {4056} (\bibinfo {year}
  {1997})}\BibitemShut {NoStop}%
\bibitem [{\citenamefont {Feynman}(1954)}]{FEY54}%
  \BibitemOpen
  \bibfield  {author} {\bibinfo {author} {\bibfnamefont {R.~P.}\ \bibnamefont
  {Feynman}},\ }\bibfield  {title} {\bibinfo {title} {Atomic theory of
  two-fluid model of liquid helium},\ }\href@noop {} {\bibfield  {journal}
  {\bibinfo  {journal} {Phys. Rev.}\ }\textbf {\bibinfo {volume} {94}},\
  \bibinfo {pages} {262} (\bibinfo {year} {1954})}\BibitemShut {NoStop}%
\bibitem [{\citenamefont {Stamper-Kurn}\ \emph {et~al.}(1999)\citenamefont
  {Stamper-Kurn}, \citenamefont {Chikkatur}, \citenamefont {Gorlitz},
  \citenamefont {Inouye}, \citenamefont {Gupta}, \citenamefont {Pritchard},\
  and\ \citenamefont {Ketterle}}]{KET99}%
  \BibitemOpen
  \bibfield  {author} {\bibinfo {author} {\bibfnamefont {D.~M.}\ \bibnamefont
  {Stamper-Kurn}}, \bibinfo {author} {\bibfnamefont {A.~P.}\ \bibnamefont
  {Chikkatur}}, \bibinfo {author} {\bibfnamefont {A.}~\bibnamefont {Gorlitz}},
  \bibinfo {author} {\bibfnamefont {S.}~\bibnamefont {Inouye}}, \bibinfo
  {author} {\bibfnamefont {S.}~\bibnamefont {Gupta}}, \bibinfo {author}
  {\bibfnamefont {D.~E.}\ \bibnamefont {Pritchard}},\ and\ \bibinfo {author}
  {\bibfnamefont {W.}~\bibnamefont {Ketterle}},\ }\bibfield  {title} {\bibinfo
  {title} {Excitations of phonons in a bose-einstein condensate by light
  scattering},\ }\href@noop {} {\bibfield  {journal} {\bibinfo  {journal}
  {Phys. Rev. Lett.}\ }\textbf {\bibinfo {volume} {83}},\ \bibinfo {pages}
  {2876} (\bibinfo {year} {1999})}\BibitemShut {NoStop}%
\bibitem [{\citenamefont {Hart}\ \emph {et~al.}(2015)\citenamefont {Hart},
  \citenamefont {Duarte}, \citenamefont {Yang}, \citenamefont {Liu},
  \citenamefont {Paiva}, \citenamefont {Khatami}, \citenamefont {Scalettar},
  \citenamefont {Trivedi}, \citenamefont {Huse},\ and\ \citenamefont
  {Hulet}}]{HUL15}%
  \BibitemOpen
  \bibfield  {author} {\bibinfo {author} {\bibfnamefont {R.~A.}\ \bibnamefont
  {Hart}}, \bibinfo {author} {\bibfnamefont {P.~M.}\ \bibnamefont {Duarte}},
  \bibinfo {author} {\bibfnamefont {T.-L.}\ \bibnamefont {Yang}}, \bibinfo
  {author} {\bibfnamefont {X.}~\bibnamefont {Liu}}, \bibinfo {author}
  {\bibfnamefont {T.}~\bibnamefont {Paiva}}, \bibinfo {author} {\bibfnamefont
  {E.}~\bibnamefont {Khatami}}, \bibinfo {author} {\bibfnamefont {R.~T.}\
  \bibnamefont {Scalettar}}, \bibinfo {author} {\bibfnamefont {N.}~\bibnamefont
  {Trivedi}}, \bibinfo {author} {\bibfnamefont {D.~A.}\ \bibnamefont {Huse}},\
  and\ \bibinfo {author} {\bibfnamefont {R.~G.}\ \bibnamefont {Hulet}},\
  }\bibfield  {title} {\bibinfo {title} {Observation of antiferromagnetic
  correlations in the hubbard model with ultracold atoms},\ }\href@noop {}
  {\bibfield  {journal} {\bibinfo  {journal} {Nature}\ }\textbf {\bibinfo
  {volume} {519}},\ \bibinfo {pages} {211} (\bibinfo {year}
  {2015})}\BibitemShut {NoStop}%
\bibitem [{\citenamefont {Mazurenko}\ \emph {et~al.}(2017)\citenamefont
  {Mazurenko}, \citenamefont {Chiu}, \citenamefont {Ji}, \citenamefont
  {Parsons}, \citenamefont {Kanasz-Nagy}, \citenamefont {Schmidt},
  \citenamefont {Grusdt}, \citenamefont {Demler}, \citenamefont {Grief},\ and\
  \citenamefont {Greiner}}]{GRE17}%
  \BibitemOpen
  \bibfield  {author} {\bibinfo {author} {\bibfnamefont {A.}~\bibnamefont
  {Mazurenko}}, \bibinfo {author} {\bibfnamefont {C.~S.}\ \bibnamefont {Chiu}},
  \bibinfo {author} {\bibfnamefont {G.}~\bibnamefont {Ji}}, \bibinfo {author}
  {\bibfnamefont {M.~F.}\ \bibnamefont {Parsons}}, \bibinfo {author}
  {\bibfnamefont {M.}~\bibnamefont {Kanasz-Nagy}}, \bibinfo {author}
  {\bibfnamefont {R.}~\bibnamefont {Schmidt}}, \bibinfo {author} {\bibfnamefont
  {F.}~\bibnamefont {Grusdt}}, \bibinfo {author} {\bibfnamefont
  {E.}~\bibnamefont {Demler}}, \bibinfo {author} {\bibfnamefont
  {D.}~\bibnamefont {Grief}},\ and\ \bibinfo {author} {\bibfnamefont
  {M.}~\bibnamefont {Greiner}},\ }\bibfield  {title} {\bibinfo {title} {A
  cold-atom fermi-hubbard antiferromagnet},\ }\href@noop {} {\bibfield
  {journal} {\bibinfo  {journal} {Nature}\ }\textbf {\bibinfo {volume} {545}},\
  \bibinfo {pages} {462} (\bibinfo {year} {2017})}\BibitemShut {NoStop}%
\bibitem [{\citenamefont {Ozeri}\ \emph {et~al.}(2005)\citenamefont {Ozeri},
  \citenamefont {Katz}, \citenamefont {Steinhauer},\ and\ \citenamefont
  {Davidson}}]{DAV05}%
  \BibitemOpen
  \bibfield  {author} {\bibinfo {author} {\bibfnamefont {R.}~\bibnamefont
  {Ozeri}}, \bibinfo {author} {\bibfnamefont {N.}~\bibnamefont {Katz}},
  \bibinfo {author} {\bibfnamefont {J.}~\bibnamefont {Steinhauer}},\ and\
  \bibinfo {author} {\bibfnamefont {N.}~\bibnamefont {Davidson}},\ }\bibfield
  {title} {\bibinfo {title} {Colloquium: Bulk bogoliubov excitations in a
  bose-einstein condensate},\ }\href@noop {} {\bibfield  {journal} {\bibinfo
  {journal} {Rev. Mod. Phys.}\ }\textbf {\bibinfo {volume} {77}},\ \bibinfo
  {pages} {187} (\bibinfo {year} {2005})}\BibitemShut {NoStop}%
\end{thebibliography}
%

\end{document}